\documentclass[a4,times]{article}
\usepackage{amsmath,amssymb,graphics,graphicx,bm,multirow,url}
\usepackage{xcolor}

\title{Modeling proportion of success in high school leaving examination- A comparative study of Inflated Unit Lindley and Inflated Beta distribution}
\author{S. Chakraborty$^1$ and S. Bhattacharjee$^{2*}$ \\
$^1$Department of Statistics, Dibrugarh University, Assam, India\\
$^2$Department of Statistics, Gauhati University, Assam, India\\
* corresponding author\\
\it Preprint} \bigskip

\begin{document}
\maketitle
\section*{Abstract}
In this article, we first introduced the inflated unit Lindley distribution considering zero or/and one inflation scenario and studied its basic distributional and structural properties. Both the distributions are shown to be members of exponential family with full rank. Different parameter estimation methods are discussed and supporting simulation studies to check their efficacy are also presented. Proportion of students passing the high school leaving examination for the schools across the state of Manipur in India for the year 2020 are then modeled using the proposed distributions  and compared with the inflated beta distribution to justify its benefits.\\

Keywords:  Inflated beta Distribution, Zero-One Inflated data, Proportion of success, MLE\\

Math Classification:\\

\section{Introduction}
In the field of applied statistics, one of the most common issues which the researchers have to deal with is data arising in terms of fractions, rates or proportions, i.e., variables which assume values in the range $\left( 0,1\right) $. However, in many cases, the data arising may contain zeroes and/or ones, i.e., the one observed in the intervals $\left[ 0,1\right) $, $\left( 0,1\right] $ or $\left[   0,1\right]   $. In such cases, continuous distributions such as Beta distribution, Kumaraswamy distribution or unit-Lindley, all of which have support in $\left( 0,1\right) $ are not suitable for modeling the data. We need probability models which are able to capture the probability mass at 0, 1 or both. Such distributions are obtained by mixing a continuous distribution having support on $\left( 0,1\right) $ with a degenerate distribution whose probability mass is concentrated at either 0 or 1 (for data arising in $\left[ 0,1\right) $ or $\left( 0,1\right] $) and with the Bernoulli distribution which assigns non-negative probability to 0 and 1 (for data arising in $\left[   0,1\right]   $) [\cite{Net19},\cite{Osp10}].\\
The idea of zero inflated continuous data modelling was first reported in \cite{Ait55} where in the authors dealt with the case of a continuous distribution which has a non-zero probability of assuming a zero value. That is, a non-zero probability mass at zero. Such distributions are mainly termed as zero inflated version of the original distribution and are obtained as a mixture of degenerate probability mass at zero with the underlying distribution. The terminology of zero inflation is common place in the case of count distributions. Alternatively the word Zero Inflated is often replaced by Zero Adjusted or Zero Spiked, etc. in the literature. Examples of occurrence of zero inflated continuous variables can be seen in many areas of application (for detail see \cite{Tom19}, \cite{Has19}, \cite{Liu19}, \cite{Bur20} and references there in).
Equally important is the situation where the observation from continuous data contains zeros and/or ones. That is, when there is situation of one-inflation or zero-one-inflation. To model such phenomenon, the probability mass at 0, at 1 or both, is included by considering mixture of continuous distribution with degenerate distribution in the case of zero or one inflation and Bernoulli in case of zero and one inflation is used. The first such was attempt seen in inflated beta distribution proposed and studied by \cite{Osp10}.\\
The main objective of this article is to consider the inflated version of the recently introduced unit Lindley distribution and briefly investigate its basic distributional properties. We consider a data in (0,1) with trace of significant zero, one inflation and analyze the data with unit Lindley and its inflated variants to establish the importance of the proposed model.\\
High school leaving examination is one of the most important milestones for the individual schools in particular and for the educational scenario of a state in general. The result of this examination declared in terms of the pass percentage of students is an important indicator, reflecting the state of affair both at the micro as well as in macro level. It is not surprising, therefore, that many of the state's education department has incentive for better show in this examination while poor show often brings penalties for the schools. It’s thus relevant to investigate the statistical modelling aspect of the school-wise pass proportion. We considered the data on result of HSLC examination from the state of Manipur in India for the academic session 2020 in this article. These data sets are available in the public domain (\cite{Boa202}, \cite{Boa201}, \cite{Boa203}) and are easily accessible. In these data sets, pass percentages are given for all the schools under the Manipur Secondary Education Board and classified with respect to three types of schools namely the, Government, Government Aided and private schools. They also provide classification of the pass results in three divisions as first division, second division and third division.\\
In this paper, the zero-or/and -one inflated unit Lindley distribution is proposed. The paper is organized as follows - Section 2 introduces the zero-or-one inflated unit Lindley distribution and some of its distributional properties, estimation of its parameters are discussed. In Section 3,  the zero-and-one inflated unit Lindley distribution is presented and some of its properties are discussed. In the next section parameter estimation  for both the distributions of preceding section is presented. Section 5 evaluates the performance of the proposed estimators through extensive simulation studies. Section 6 contains  empirical applications of the proposed inflated unit Lindley distribution in comparison to the inflated beta distrbution for four data sets. The paper is concluded with some final remarks presented in Section 7.
\section{The Zero-or-One inflated unit Lindley distribution}
\begin{figure}[h!]
	\includegraphics[keepaspectratio,height=13cm,width=13cm]{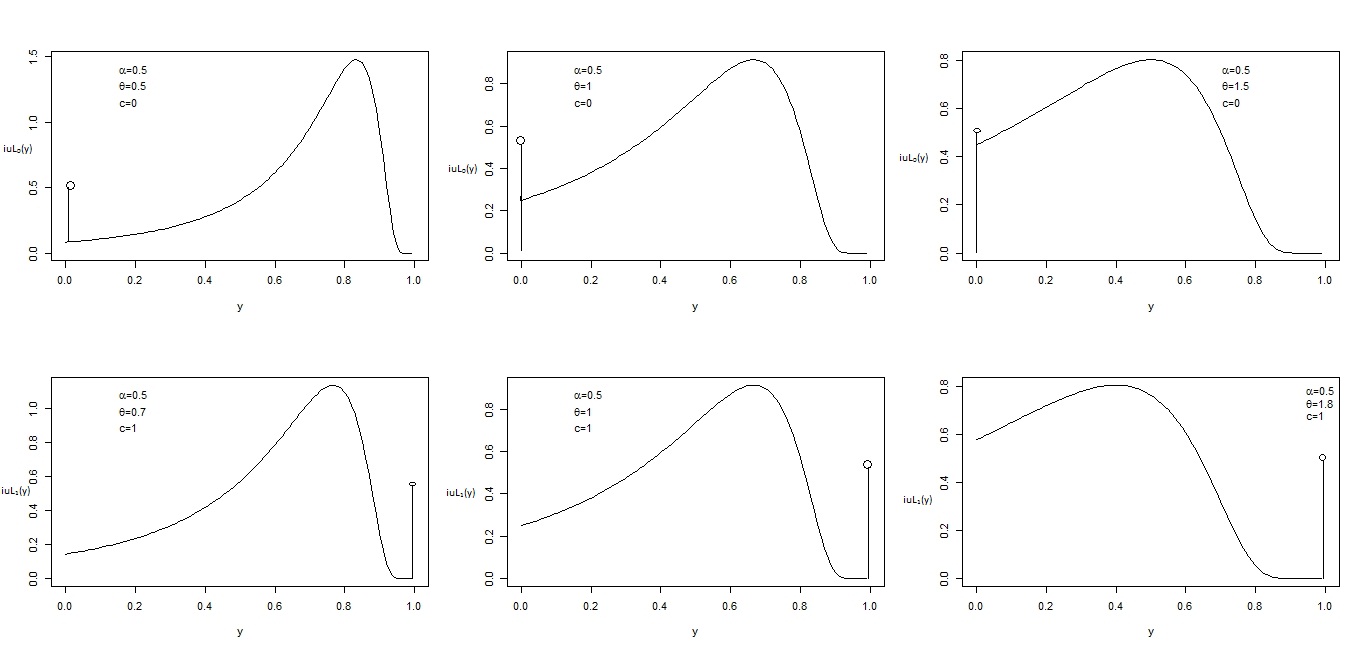}
	\caption{ULZI and ULOI densities for different values of $\theta$ and $c$; $\alpha=0.5$}\label{fig1}
\end{figure}

The unit Lindley distribution is a one parameter continuous distribution having support on $\left( 0,1\right) $ which is obtained from the Lindley distribution through a transformation \cite{Maz19}. It has certain advantages over the commonly used beta distribution with two parameters defined in the range $\left( 0,1\right) $ \cite{John95} such as closed form of c.d.f., quantile function and simple expressions for moments. It scores over the competing Kumaraswamy distribution with respect to the fact that there is no closed form of the moments of this distribution. This distribution also enabled the development of a new bounded regression model which is a feasible alternative to the popular Beta regression model. The unit Lindley distribution with parameter $\theta$ has the p.d.f.
\begin{equation}\label{eq1}
f\left( x|\theta\right) =\dfrac{\theta^2}{1+\theta}\left( 1-x\right) ^{-3}\exp\left( -\dfrac{\theta x}{1-x}\right) ,\hspace{0.5cm}\theta>0,x \in \left( 0,1\right) 
\end{equation}
The real-life data may include values such as zeroes and/or ones. In such cases, one needs to focus on incorporating a discrete component into the continuous data generating process so that the values zeroes and/or ones are observed with a positive probability. We, thus, consider a mixture of two distributions: the continuous unit Lindley distribution on $\left( 0,1\right) $ and a degenerate distribution having the entire probability mass concentrated at the known point $c$, where $c=0$ or $c=1$. We refer to it as the data being inflated (having higher probability of occurrence) at one/both endpoints of the standard unit interval.
The c.d.f. of the mixture distribution, known as the Inflated unit Lindley distribution is given by
\begin{equation}\label{eq2}
IuL_c\left( y;\alpha,\theta\right) =\alpha\mathbb{I}_{\left[ c,1\right] }\left( y\right) +\left( 1-\alpha\right) F\left( y;\theta\right) 
\end{equation}
where $\mathbb{I}_A$ is the indicator function which takes the value 0 if $y \in A$ and 1 if $y \notin A$, $0<\alpha<1$ is the mixture parameter and $F\left( ;\theta\right) $ is the c.d.f. of the unit Lindley distribution with parameter $\theta$. Here, the r.v. $Y$ follows the unit Lindley distribution with parameter $\theta$ with probability $\left( 1-\alpha\right)$ and the degenerate distribution at $c$ with probability $\alpha$. The p.d.f. of the Inflated unit Lindley distribution corresponding to the c.d.f in (\ref{eq2}) is given by
\begin{equation}\label{eq3}
iuL_c \left( y;\alpha,\theta\right) =\begin{cases}
\alpha, \hspace{2.05cm}\text{if   } y=c\\
\left( 1-\alpha\right) f\left( y;\theta\right), \text{if   } y \in \left( 0,1\right)
\end{cases}
\end{equation}
where $f\left( ;\theta\right)$ is the unit Lindley density in (\ref{eq1}) and $\alpha \in \left( 0,1\right)$ is the probability mass at $c$, representing the probability of observing 0 (when $c=0$) or 1 (when $c=1$).\\
\noindent \textbf{Definition 2.1} Let $Y$ be a r.v. that follows the inflated unit Lindley distribution in (\ref{eq3}). 
\begin{enumerate}
	\item If $c=0$, the distribution in (\ref{eq3}) is called the zero-inflated unit Lindley distribution (ULZI) and we write $Y \sim ULZI\left( \alpha,\theta\right) $, where $\alpha=P\left( Y=0\right) $.
	\item If $c=1$, the distribution in (\ref{eq3}) is called the one-inflated unit Lindley distribution (ULOI) and we write $Y \sim ULOI\left( \alpha,\theta\right) $, where $\alpha=P\left( Y=1\right) $.
\end{enumerate}
The $r^{th}$ raw moment of $Y$ is
\begin{equation*}
E\left( Y^r\right) =\alpha c+\left( 1-\alpha\right) \mu_r^{\prime}; r=1,2,...
\end{equation*}
where $\mu_r^{\prime}=\dfrac{r}{1+\theta}\int_{0}^{1}\left[ \dfrac{x^{r-1}\left( 1-\theta+x\right) }{\left( 1-x\right) }\exp\left( -\dfrac{\theta x}{1-x}\right)\right] dx $ is the $r^{th}$ raw moment of the unit Lindley distribution. In particular, the mean and variance of $Y$ are
\begin{equation}\label{eqn4}
E\left( Y\right) =\alpha c+\left( 1-\alpha\right) \mu_1^{\prime}=\alpha c+\dfrac{\left( 1-\alpha\right)}{\left( 1+\theta\right)}
\end{equation}
\begin{equation}\label{eqn5}
Var\left( Y\right)=\alpha c\left( 1-\alpha c\right) +\dfrac{1-\alpha}{\left(   1+\theta\right) }\left[ \left\lbrace \theta^2e^\theta Ei\left( 1,\theta\right)  -\theta+1\right\rbrace -2\alpha c -\dfrac{1-\alpha}{\left( 1+\theta\right) }\right]
\end{equation}
where $Ei\left( 1,\theta\right)$ represents the exponential integral function \cite{Maz19}.\\
Figure (\ref{fig1}) shows the ULZI and ULOI densities inflated at $c=0$ and $c=1$ for different values of $\theta$ and with the value of the mixing parameter $\alpha$ fixed at 0.5. It can be observed that the p.d.f. of both the ULZI and ULZO distribution assume different shapes for different values of $\theta$ and $c$. $\theta$ is seen to control the shape of the probability curve, whose skewness increases with an increase in the value of $\theta$. In all the sub plots in figure (\ref{fig1}), the vertical bar with a circle above represents $\alpha=0.5$ ($P\left( Y=0\right) $ or $P\left( Y=1\right) $). Further, both the ULZI and ULOI distributions have the same functional shape in $\left( 0,1\right)$, the only difference being in the mass point, which is 0 for the ULZI and 1 for the ULOI distribution.\\
\newpage
\noindent \textbf{Result1} The zero- and one-inflated unit Lindley distribution in (\ref{eq3}) are  member of two-parameter 	exponential family distributions of full rank.\\
\noindent \textbf{Proof:} 
Denoting
\begin{eqnarray*}
T_1(y)&=&I_{c}(y)=\begin{cases}
1, \hspace{2.05cm}\text{if   } y=c\\
0, \text{if   } y \in \left( 0,1\right)
\end{cases}\\
T_2(y)&=&\begin{cases}
1, \hspace{2.05cm}\text{if   } y=c\\
0, \text{if   } y \in \left( 0,1\right)
\end{cases}\\
\eta_1&=&\log\frac{\alpha}{1-\alpha}+\log\frac{\theta^2}{1+\theta}\\
\eta_2&=&\theta\\
\end{eqnarray*}
We can rewrite the p.d.f. of the zero- and one-inflated unit Lindley distribution given in the equation (\ref{eq3}) as 
\begin{eqnarray*}
iuL_c\left( y;\alpha,\theta\right) &=&(1-y\mathbb{I}_{\left[ c,1\right] }\left( y\right))^{-3} \exp[ \left( \log(\frac{\alpha}{1-\alpha}+\log\frac{1+\theta}{\theta^2}\right)T_1(y)+\eta_2 T_2(y)\\
&&\-\log\left\{1+\exp\left(\eta_1-\log\frac{1+\theta}{\theta^2}\right)\right\}-\log\frac{1+\theta}{\theta^2}] 
\end{eqnarray*}
Now taking
\begin{eqnarray*}
T(y)&=&\left(T_1(y), T_2(y)\right),\\
\eta&=&\left(\eta_1, \eta_2\right),\\
B^*(\eta)&=&\log\left\{1+\exp\left(\eta_1-B(\eta_2)\right)\right\},\\
B(\eta_2)&=&\log\frac{1+\eta_2}{\eta_2^2}\;\;\text{and}\\
h(y)&=&(1-y\mathbb{I}_{\left[ c,1\right] }\left( y\right))^{-3}\theta
\end{eqnarray*} 
the p.d.f. can finally be expressed as 
$$\exp\left[\eta^{\prime}T(y)-B^*(\eta)\right]h(y).$$ Note that the functions $B^*(\eta)$ is a real valued function of $\eta_1, \eta_2$,$h(y)$ is a positive real valued function. The transformation from $(\alpha, \theta)$ to $(\eta_1,  \eta_2)$ is obviously one-one from $(0, 1) \times R^+$ to $R \times R^+$. Hence the p.d.f. in (\ref{eq3}) belongs to a two parameter exponential family distribution of full rank.\\

\section{The Zero-and-One-inflated unit Lindley distribution} 
\begin{figure}[h!]
	\includegraphics[keepaspectratio,height=13cm,width=13cm]{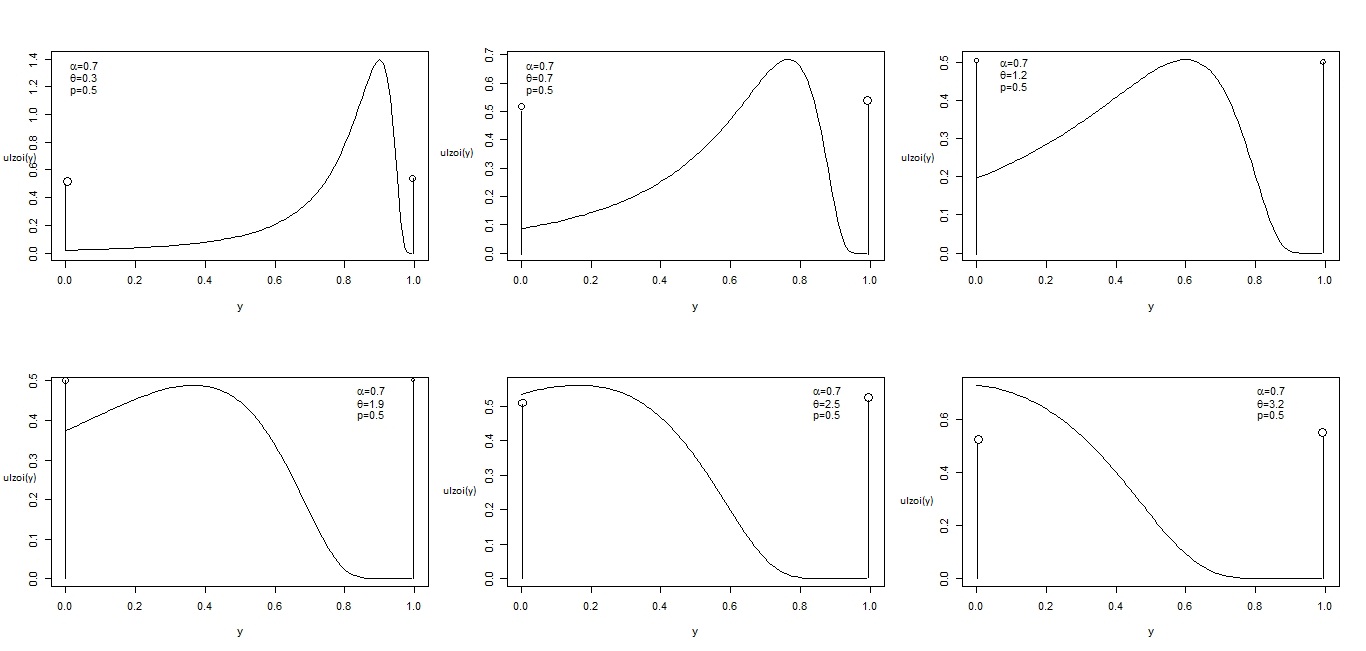}
	\caption{ULZOI density for different values of $\theta$; $\alpha=0.7~\text{and}~p=0.5$}\label{fig2}
\end{figure}

The Zero-or-One inflated unit Lindley distribution introduced in the previous section is suitable for modeling data which has data inflation on either of the two end points of the standard unit interval $\left( 0,1\right) ,$ but not on both the end points. To model data arising in the interval $\left[ 0,1\right] $, we need a mixture of the unit Lindley distribution and the Bernoulli distribution, which assigns non-negative probability to the points 0 and 1. The c.d.f. of the mixture distribution, known as the Zero-and-One-inflated unit Lindley distribution (ULZOI) is given by
\begin{equation}\label{eqn6}
ULZOI\left( y;\alpha,p,\theta\right) =\alpha Ber\left( y;p\right) + \left( 1-\alpha\right) F\left( y;\theta\right) 
\end{equation} 
where $y \in \left[ 0,1\right] $, $Ber\left(;p\right) $ represents the c.d.f. of a Bernoulli r.v. and $F\left( ;\theta\right) $ is the c.d.f. of the unit Lindley distribution with parameter $\theta$. Further, $\alpha$ is the mixing parameter which lies between 0 and 1.
We say that a r.v. $Y$ assuming values in $\left[ 0,1\right] $ has a ULZOI distribution with parameters $\alpha$, $p$ and $\theta$ if its density function with respect to the c.d.f. in (\ref{eqn6}) is given by  
\begin{equation}\label{eqn7}
ulzoi\left( y;\alpha,p,\theta\right) =\begin{cases}
\alpha p,\hspace{3.05cm}\text{if   } y=1\\
\alpha \left(1-p\right), \hspace{2.05cm}\text{if   } y=0\\
\left( 1-\alpha\right) f\left( y;\theta\right),\hspace{1.15cm} \text{if   } y \in \left( 0,1\right)\\
\end{cases}
\end{equation}
where $0<\alpha,p<1$, $,\theta>0$ and $f\left( y;\theta\right)$ is the p.d.f. of the unit Lindley distribution in (\ref{eq1}). We write $Y \sim ULZOI(\alpha,p,\theta)$ and in that case, $\alpha p = P(Y=1)$ and $\alpha \left(1-p\right)=P(Y=0)$. \\
Also, the $r^{th}$ raw moment of $Y$ is
\begin{equation*}
E\left( Y^r\right) =\alpha p+\left( 1-\alpha\right) \mu_r^{\prime}; r=1,2,...
\end{equation*}
Consequently, the mean and variance of $Y$ are
\begin{equation}\label{eqn8}
E\left( Y\right) =\alpha p+\left( 1-\alpha\right) \mu_1^{\prime}=\alpha p+\dfrac{\left( 1-\alpha\right)}{\left( 1+\theta\right)}
\end{equation}
\begin{equation}\label{eqn9}
Var\left( Y\right)=\alpha p\left( 1-\alpha p\right) +\dfrac{1-\alpha}{\left(   1+\theta\right) }\left[ \left\lbrace \theta^2e^\theta Ei\left( 1,\theta\right)  -\theta+1\right\rbrace -2\alpha p -\dfrac{1-\alpha}{\left( 1+\theta\right) }\right] 
\end{equation}
where $Ei\left( 1,\theta\right)$ represents the exponential integral function.
Figure (\ref{fig2}) presents the ULZOI density for different values of $\theta$ for $\alpha=0.7$ and $p=0.5$. It is evident from this plot that as the value of $\theta$ increases, the skewness also increases and also, for higher values of $\theta$, the probability curve of ULZOI distribution approaches the reverse sigmoid curve. 
\\

\noindent \textbf{Result2} The zero-and-one-inflated unit Lindley distribution in (\ref{eq3}) is a three parameter
exponential family distribution of full rank.\\
\noindent \textbf{Proof:} 
Denoting
\begin{eqnarray*}
T_1(y)&=&I_{\{0,1\}}(y)=\begin{cases}
1, \hspace{2.05cm}\text{if   } y=0\,\, \text{or}\,\, 1\\
0,  \hspace{2.05cm}\text{if   } y \in \left( 0,1\right)
\end{cases}\\
T_2(y)&=&yI_{\{0,1\}}(y)=\begin{cases}
1, \hspace{2.05cm}\text{if   } y=1\\
0,  \hspace{2.05cm}\text{if   } y=0
\end{cases}\\
T_3(y)&=&\dfrac{y}{1-y}\\
\eta_1&=&\log\frac{\alpha}{1-\alpha}+\log\frac{1+\theta}{\theta^2}-\log\frac{1}{1-p}\\
\eta_2&=&\log\frac{1}{1-p}\\
\eta_3&=&\theta\\
\end{eqnarray*}
We can rewrite the p.d.f. of the zero- and one-inflated unit Lindley distribution given in the equation (\ref{eqn7}) as 
\begin{eqnarray*}
&&ulzoi\left( y;\alpha,p,\theta\right)= (1-y\mathbb{I}_{\left( 0,1\right) }\left( y\right))^{-3} \exp[\left( \log(\frac{\alpha}{1-\alpha}+\log\frac{1+\theta}{\theta^2}-\log\frac{1}{1-p}\right)T_1(y)\\&+&\log\frac{1}{1-p}T_2(y)+\theta T_3(y)-\log\left\{1+\exp\left(\eta_1-\log\frac{1+\theta}{\theta^2}+\log\frac{1}{1-p}\right)\right\}-\log\frac{1+\theta}{\theta^2}]
\end{eqnarray*}
Now taking
\begin{eqnarray*}
T(y)&=&\left(T_1(y), T_2(y), T_3(y)\right),\\
\eta&=&\left(\eta_1, \eta_2, \eta_3\right),\\
B^*(\eta)&=&\log\left\{1+\exp\left(\eta_1+M(\eta_2)-N(\eta_3)\right)\right\},\\
M(\eta_2)&=&\log(1+e^{\eta_2})\\
N(\eta_3)&=&\log\frac{1+\eta_3}{\eta_3^2}\;\;\text{and}\\
h(y)&=&(1-y\mathbb{I}_{\left[ 0,1\right] }\left( y\right))^{-3}\theta
\end{eqnarray*}
 
the p.d.f. can finally be expressed as 
$$\exp\left[\eta^{\prime}T(y)-B^*(\eta)\right]h(y).$$ Note that the functions $B^*(\eta)$ is a real valued function of $\left(\eta_1, \eta_2, \eta_3\right)$,$h(y)$ is a positive real valued function. The transformation from $(\alpha, p, \theta)$ to $(\eta_1,  \eta_2, \eta_3)$ is obviously one-one from $(0, 1) \times (0, 1)\times R^+$ to $R\times R^+ \times R^+$. Also neither the $t$s nor the $\eta$s are linearly related. Hence the p.d.f. in (\ref{eqn7}) belongs to a three parameter exponential family distribution of full rank.\\

\section{Estimation and related} 
In this section, the maximum likelihood estimation (MLE) of the parameters and construction of the Fisher Information Matrix is considered.
\subsection{MLE: Zero or One inflated ULD}
The likelihood function for $\boldsymbol{\nu}=\left( \alpha,\theta\right) $ based on the $iuL_c$ random sample $\boldsymbol{y}=\left( y_1,y_2,\ldots,y_n\right) ^\prime$ is given by
\begin{eqnarray*}
L\left( \boldsymbol{\nu},\boldsymbol{y}\right) &=&\prod_{i=1}^{n}iuL_c\left( y_i;\alpha,\theta\right) \\
&=&\prod_{i=1}^{n}\left[ \left\lbrace \alpha^{\mathbb{I}_{\left[ c\right]} \left( y_i\right)}\left( 1-\alpha\right) ^{1-\mathbb{I}_{\left[ c\right]}\left( y_i\right)} \right\rbrace \left\lbrace f\left( y,\theta\right) ^{1-\mathbb{I}_{\left[ c\right]}\left( y_i\right)} \right\rbrace \right] \\
&=& L_1\left( \alpha,y\right) \times L_2\left( \theta,y\right)
\end{eqnarray*}
It can be seen that the likelihood function is factorized into two terms $L_1$ and $L_2$ where $L_1$ depends on $\alpha$ and $L_2$ only on $\theta$. Now,
\begin{eqnarray*}
L_1\left( \alpha,y\right)&=&\prod_{i=1}^{n}\alpha^{\mathbb{I}_{\left[ c\right]} \left( y_i\right)}\left( 1-\alpha\right) ^{1-\mathbb{I}_{\left[ c\right]}\left( y_i\right)}\\
&=& \alpha^{\sum_{i=1}^{n} \mathbb{I}_{\left[ c\right]} \left( y_i\right)}\left( 1-\alpha\right) ^{n-\sum_{i=1}^{n} \mathbb{I}_{\left[ c\right]} \left( y_i\right)}
\end{eqnarray*}
and
\begin{eqnarray*}
L_2\left(  \theta,y\right) &=& \underset{y_i \in \left( 0,1\right) }{\prod_{i=1}^{n}}f\left( y_i;\theta\right) ^{1-\mathbb{I}_{\left[ c\right]}\left( y_i\right)}
\end{eqnarray*}
The log-likelihood function of the Zero-or-One Inflated Unit Lindley distribution is given by
$l\left( \boldsymbol{\nu}\right) =\log L\left( \boldsymbol{\nu},\boldsymbol{y}\right)=l_1\left( \alpha;y\right) +l_2\left( \theta;y\right) $ 
where
\begin{eqnarray*}
l_1\left( \alpha;y\right)&=&\log L_1\left( \alpha,y\right)\\
&=& \log\alpha \sum_{i=1}^{n}\mathbb{I}_{\left[ c\right]}\left( y_i\right)+\log\left( 1-\alpha\right) \left\lbrace {n-\sum_{i=1}^{n} \mathbb{I}_{\left[ c\right]} \left( y_i\right)}\right\rbrace 
\end{eqnarray*}
and
\begin{eqnarray*}
l_2\left( \theta;y\right)&=&\log L_2\left( \theta,y\right)\\
&=& \left\lbrace n-{\sum_{i=1}^{n}} \mathbb{I}_{\left[ c\right] }\left( y_i\right)\right\rbrace\log \left( \dfrac{\theta^2}{1+\theta}\right) -3\underset{y_i \in \left( 0,1\right) }{\sum_{i=1}^{n}}\log \left( 1-y_i\right) -\theta \underset{y_i \in \left( 0,1\right) }{\sum_{i=1}^{n}}\dfrac{y_i}{1-y_i}
\end{eqnarray*}
The score function is then obtained by differentiating the log-likelihood function and is denoted by $U\left( \boldsymbol{\nu}\right) =\left[ U_\alpha\left( \alpha\right) ,U_\theta\left( \theta\right)\right]  $, where
\begin{eqnarray*}
U_\alpha\left( \alpha\right) &=&\dfrac{\partial l_1\left( \alpha;y\right)}{\partial \alpha}\\
&=&\dfrac{1}{\alpha}\sum_{i=1}^{n}\mathbb{I}_{\left[ c\right]}\left( y_i\right)-\dfrac{1}{1-\alpha}\left\lbrace {n-\sum_{i=1}^{n} \mathbb{I}_{\left[ c\right]} \left( y_i\right)}\right\rbrace\\
U_\theta\left( \theta\right) &=&\dfrac{\partial l_2\left( \theta;y\right)}{\partial \theta} \\
&=& \dfrac{2+\theta}{\theta\left( 1+\theta\right) }\left\lbrace n-{\sum_{i=1}^{n}} \mathbb{I}_{\left[ c\right] }\left( y_i\right)\right\rbrace-\underset{y_i \in \left( 0,1\right) }{\sum_{i=1}^{n}}\dfrac{y_i}{1-y_i}
\end{eqnarray*}
The maximum likelihood estimator of $\alpha$ is $\hat{\alpha}=\dfrac{1}{n}\sum_{i=1}^{n}\mathbb{I}_{\left[ c\right] }\left( y_i\right)$ , i.e., the proportion of $n$ values that are equal to $c$ and the maximum likelihood estimator of $\theta$ is 
\begin{eqnarray*}
	&&\hat{\theta}=\frac{1}{2\underset{y_i \in \left( 0,1\right) }{\sum_{i=1}^{n}}\dfrac{y_i}{1-y_i}}}{\left( n-\underset{y_i \in \left( 0,1\right) }{\sum_{i=1}^{n}} \mathbb{I}_{\left[ c\right] }\left( y_i\right)
	-\underset{y_i \in \left( 0,1\right) }{\sum_{i=1}^{n}}\dfrac{y_i}{1-y_i}+\right.\\
	&&\left. \left[ \left( \underset{y_i \in \left( 0,1\right) }{\sum_{i=1}^{n}}\dfrac{y_i}{1-y_i}\right)^2+\left\lbrace n- \sum_{i=1}^{n}  \mathbb{I}_{\left[ c\right] }\left( y_i\right)\right\rbrace  ^2-
	6\underset{y_i \in \left( 0,1\right) }{\sum_{i=1}^{n}}\dfrac{y_i}{1-y_i}\left(n- \sum_{i=1}^{n}  \mathbb{I}_{\left[ c\right] }\left( y_i\right)\right)  \right] ^\frac{1}{2}\right)
\end{eqnarray*}
The Fisher Information matrix for the Zero or One inflated Unit Lindley law is
$$
	K\left( \boldsymbol{\nu}\right) =
\begin{bmatrix}
k_{\alpha\alpha} & k_{\alpha\theta}\\
k_{\theta\alpha} & k_{\theta\theta}\\
\end{bmatrix}
\quad
$$
where
\begin{eqnarray*}
k_{\alpha\alpha}&=&-E\left[ \dfrac{\partial U_\alpha\left( \alpha\right) }{\partial \alpha}\right] \\
&=& \dfrac{n}{\alpha\left( 1-\alpha\right) }\\
k_{\alpha\theta}&=&-E\left[ \dfrac{\partial U_\theta\left( \theta\right) }{\partial \alpha}\right]\\
&=&0\\
k_{\theta\alpha}&=& -E\left[ \dfrac{\partial U_\alpha\left( \alpha\right) }{\partial \theta}\right] \\
&=&0\\
k_{\theta\theta}&=&  -E\left[ \dfrac{\partial U_\theta\left( \theta\right) }{\partial \theta}\right] \\
&=& \dfrac{n\left( 1-\alpha\right) \left( \theta^2+4\theta+2\right) }{\theta^2\left( 1+\theta\right) ^2}
\end{eqnarray*}
Suppose $\hat{\nu}=\left( \hat{\alpha},\hat{\theta}\right) $ denote the m.l.e. of $\boldsymbol{\nu}=\left( \alpha,\theta\right) $. In large samples, $\hat{\nu}$ is asymptotically normally distributed, i.e., $\hat{\nu}\xrightarrow{D}N_2\left( \nu,k\left( \nu\right) ^{-1}\right) $ where $k\left( \nu\right) $ is the Fisher Information Matrix. Using this result, approximate confidence intervals for the parameters $\alpha$ and $\theta$ can be constructed. Let $\delta \in \left( 0,0.5\right) $. Then $\left( 1-\delta\right) \times 100\%$ asymptotic confidence intervals for $\alpha$ and $\theta$ are given respectively by\\
\begin{center}
$\hat{\alpha}\pm Z_{1-\frac{\alpha}{2}}SE\left( \hat{\alpha}\right) $ and $\hat{\theta}\pm Z_{1-\frac{\alpha}{2}}SE\left( \hat{\theta}\right) $
\end{center}
where $SE\left( .\right) $ denotes standard error and $Z_{1-\frac{\alpha}{2}}$ is the standard normal quantile.
\medskip
\subsection{MLE: Zero and One inflated ULD}
The p.d.f. of the Zero and One inflated unit Lindley distribution can be re-written as
\begin{eqnarray*}
ulzoi\left( y;\alpha,p,\theta\right)&=&\left[\alpha p^y \left( 1-p\right)^{1-y} \right]^ {\mathbb{I}_{\left\{ 0,1 \right\} }\left( y\right)}\left[ \left( 1-\alpha\right) f\left( y;\theta\right) \right]^{1-\mathbb{I}_{\left\{ 0,1 \right\} }\left( y\right)} \\
&=&\left[ \alpha^{\mathbb{I}_{\left[ 0,1\right] }\left( y\right)}\left( 1-\alpha\right) ^{1-\mathbb{I}_{\left\{ 0,1 \right\} }\left( y\right)}\right] \left[ p^y \left( 1-p\right)^{1-y} \right]^ {\mathbb{I}_{\left\{ 0,1\right\} }\left( y\right)}\\
&&\left[ f\left( y;\theta\right)\right] ^{1-\mathbb{I}_{\left\{ 0,1\right\} }\left( y\right)}
\end{eqnarray*}
where $\mathbb{I}_{\left\{ 0,1\right\} }\left( y\right)$ is an indicator function such that\\
$
\mathbb{I}_{\left\{ 0,1\right\} }\left( y\right)=
\begin{cases}
	1,\text{if}~y\in \left\{ 0,1\right\} \\
	0, \text{if}~y \notin  \left\{ 0,1\right\}
\end{cases}
$
\\
Here, it can observed that the first term depends only on $\alpha$, the second term depends only on $p$ and the third term depends only on $\theta$. The likelihood function for $\boldsymbol{\nu}=\left( \alpha,p,\theta\right) $ on the random sample $\left( y_1,y_2,\ldots,y_n\right) $ from the $ulzoi\left( y;\alpha,p,\theta\right)$ distribution is given by
\begin{eqnarray*}
L\left( \boldsymbol{\nu},\boldsymbol{y}\right) &=& \prod_{i=1}^{n}ulzoi\left( y_i;\alpha,p,\theta \right) \\
&=& L_1\left( \alpha;y\right) \times L_2\left( p;y\right) \times L_3\left( \theta;y\right)  
\end{eqnarray*}
where
\begin{eqnarray*}
L_1\left( \alpha;y\right) &=& \prod_{i=1}^{n}\left[ \alpha^{\mathbb{I}_{\left\{ 0,1\right\} }\left( y_i\right)}\left( 1-\alpha\right) ^{1-\mathbb{I}_{\left\{ 0,1\right\} }\left( y_i\right)}\right]\\
&=& \alpha^{\sum_{i=1}^{n}\mathbb{I}_{\left\{ 0,1\right\} }\left( y_i\right)}\left( 1-\alpha\right) ^{n-\mathbb{I}_{\left\{ 0,1\right\} }\left( y_i\right)}\\
L_2\left( p;y\right) &=& \prod_{i=1}^{n}\left[ p^{y_i} \left( 1-p\right)^{1-y_i} \right]^ {\mathbb{I}_{\left\{ 0,1\right\} }\left( y_i\right)}\\
&=& p^{\sum_{i=1}^{n}y_i \mathbb{I}_{\left\{ 0, 1\right\} }\left( y_i\right)}\left( 1-p\right) ^{\sum_{i=1}^{n}\left( 1-y_i\right)  \mathbb{I}_{\left\{ 0,1\right\} }\left( y_i\right)}\\
L_3\left( \theta;y\right) &=& \underset{y_i\in\left( 0,1\right) }{\prod_{i=1}^{n}}f\left( y_i;\theta\right)\\
&=& \exp\left( -\theta \sum_{i=1}^{n}\dfrac{y_i}{1-y_i}\right) \prod_{i=1}^{n} \left( \dfrac{\theta^2}{1+\theta}\right)\left( 1-y_i\right) ^{-3} 
\end{eqnarray*}
The corresponding log-likelihood function is
$l\left( \boldsymbol{\nu}\right) =\log L\left( \boldsymbol{\nu},\boldsymbol{y}\right) = l_1\left( \alpha;y\right)+l_2\left( p;y\right)+l_3\left( \theta;y\right)$ 
where
\begin{eqnarray*}
l_1\left( \alpha;y\right) &=&\log \alpha\sum_{i=1}^{n}\mathbb{I}_{\left\{ 0,1\right\} }\left( y_i\right)+\log\left( 1-\alpha\right) \left\lbrace n-\sum_{i=1}^{n}\mathbb{I}_{\left\{0, 1\right\} }\left( y_i\right)\right\rbrace \\
l_2\left( p;y\right) &=&\log p \sum_{i=1}^{n} y_i \mathbb{I}_{\left\{0,  1\right\} }\left( y_i\right)+\log\left( 1-p\right) \left[ \sum_{i=1}^{n}\mathbb{I}_{\left\{ 0,1\right\} }\left( y_i\right)-\sum_{i=1}^{n}y_i \mathbb{I}_{\left\{ 0, 1\right\} }\left( y_i\right)\right] \\
l_3\left( \theta;y\right) &=& \underset{y_i\in\left( 0,1\right) }{\sum_{i=1}^{n}}\log \left( \dfrac{\theta^2}{1+\theta}\right) -3\underset{y_i\in\left( 0,1\right) }{\sum_{i=1}^{n}}\log \left( 1-y_i\right) -\theta \underset{y_i\in\left( 0,1\right) }{\sum_{i=1}^{n}}\dfrac{y_i}{1-y_i}
\end{eqnarray*}
The score function is denoted by
\begin{center}
	$U\left( \boldsymbol{\nu}\right) =\left[ U_\alpha\left( \alpha\right) ,U_p\left( p\right) ,U_\theta\left( \theta\right)\right]  $
\end{center}
where
\begin{eqnarray*}
U_\alpha\left( \alpha\right)&=&\dfrac{\partial l_1\left( \alpha;y\right)}{\partial \alpha}\\
&=&\dfrac{1}{\alpha}\sum_{i=1}^{n}\mathbb{I}_{\left\{ 0,1\right\} }\left( y_i\right)-\dfrac{1}{1-\alpha}\left[n-\mathbb{I}_{\left\{ 0,1\right\} }\left( y_i\right) \right] \\
U_p\left( p\right)&=&  \dfrac{1}{p}\sum_{i=1}^{n} y_i \mathbb{I}_{\left\{0, 1\right\} }\left( y_i\right)-\dfrac{1}{1-p}\left[\mathbb{I}_{\left\{ 0,1\right\} }\left( y_i\right) -\mathbb{I}_{\left[ 1\right] }\left( y_i\right)\right] \\
U_\theta\left( \theta\right) &=& \dfrac{2+\theta}{\theta\left( 1+\theta\right) }\left[n-{\sum_{i=1}^{n}}\mathbb{I}_{\left\{ 0,1\right\} }\left( y_i\right) \right]-\underset{y_i\in \left( 0,1\right)}{\sum_{i=1}^{n}}\dfrac{y_i}{1-y_i}
\end{eqnarray*}
The m.l.e of $\alpha$ is $\hat{\alpha}=\dfrac{1}{n}\underset{y_i\in \left( 0,1\right) }{\sum_{i=1}^{n}}\mathbb{I}_{\left\{ 0,1\right\} }\left( y_i\right)$, which is the proportion of discrete values in the sample. The m.l.e of $p$ is $\hat{p}=\dfrac{\sum_{i=1}^{n}\mathbb{I}_{\left\{ 1\right\} }\left( y_i\right)}{{\sum_{i=1}^{n}}\mathbb{I}_{\left\{ 0,1\right\} }\left( y_i\right)}$. The m.l.e of $\theta$ is\\
\begin{eqnarray*}
	&&\hat{\theta}=\frac{1}{2\underset{y_i \in \left( 0,1\right) }{\sum_{i=1}^{n}}\dfrac{y_i}{1-y_i}}}{\left( n-\underset{y_i \in \left( 0,1\right) }{\sum_{i=1}^{n}} \mathbb{I}_{\left[ c\right] }\left( y_i\right)
	-\underset{y_i \in \left( 0,1\right) }{\sum_{i=1}^{n}}\dfrac{y_i}{1-y_i}+\right.\\
	&&\left. \left[ \left( \underset{y_i \in \left( 0,1\right) }{\sum_{i=1}^{n}}\dfrac{y_i}{1-y_i}\right)^2+\left\lbrace n- \sum_{i=1}^{n}  \mathbb{I}_{\left[ c\right] }\left( y_i\right)\right\rbrace  ^2-
	6\underset{y_i \in \left( 0,1\right) }{\sum_{i=1}^{n}}\dfrac{y_i}{1-y_i}\left(n- \sum_{i=1}^{n}  \mathbb{I}_{\left[ c\right] }\left( y_i\right)\right)  \right] ^\frac{1}{2}\right)
\end{eqnarray*}
The Fisher Information matrix for the Zero and One inflated Unit Lindley law is
$$
K\left( \boldsymbol{\nu}\right) =
\begin{bmatrix}
k_{\alpha\alpha} & k_{\alpha p} & k_{\alpha\theta}\\
k_{p\alpha} & k_{pp} & k_{p\theta}\\
k_{\theta\alpha} & k_{\theta p} & k_{\theta\theta}\\
\end{bmatrix}
\quad
$$
where
$	k_{\alpha\alpha}=\dfrac{n}{\alpha\left( 1-\alpha\right) }$\\
	$k_{\alpha p}=k_{\alpha\theta}=k_{p\alpha}=k_{p\theta}=k_{\theta\alpha}=k_{\theta p}=0$\\
$	k_{pp}= \dfrac{n\alpha}{p\left( 1-p\right) } $\\
	$k_{\theta\theta}=\dfrac{n\left( 1-\alpha\right) \left( \theta^2+4\theta+2\right) }{\theta^2\left( 1+\theta\right) ^2} $\\
Suppose $\hat{\nu}=\left( \hat{\alpha},\hat{p},\hat{\theta}\right) $. In large samples, $\hat{\nu}$ is asymptotically normally distributed, i.e., $\hat{\nu}\xrightarrow{D}N_3\left( \nu,k\left( \nu\right) ^{-1}\right) $ where $k\left( \nu\right) $ is the Fisher Information Matrix. Let $\delta \in \left( 0,0.5\right) $. Then $\left( 1-\delta\right) \times 100\%$ asymptotic confidence intervals for $\alpha$, $p$ and $\theta$ are given respectively by\\
\begin{center}
	$\hat{\alpha}\pm Z_{1-\frac{\alpha}{2}}SE\left( \hat{\alpha}\right) $, $\hat{p}\pm Z_{1-\frac{\alpha}{2}}SE\left( \hat{p}\right)$ and $\hat{\theta}\pm Z_{1-\frac{\alpha}{2}}SE\left( \hat{\theta}\right) $
\end{center}
where the symbols have their usual meaning.

\subsection{Conditional Mean Estimation}
The conditional mean of $Y \sim ULZI\left( \alpha,\theta\right) $ given $y\in \left( 0,1\right) $ is obtained as:
\begin{eqnarray}
	\nonumber E\left( Y|y\in \left( 0,1\right)\right) &=& \int_{0}^{1}y iuL_c\left( y;\alpha,\theta\right)dy\\\nonumber
	&=& \dfrac{1}{1+\theta}
\end{eqnarray}
which does not depend on $\alpha$. Consequently, the conditional mean estimate (CME) of $\theta$ is $\left\lbrace \dfrac{1}{E\left( Y|y\in \left( 0,1\right)\right)}-1\right\rbrace $. The CME of $ \theta $ for $Y\sim ULZOI\left( y;\alpha,p,\theta\right)$ is also obtained likewise.

\section{Assessment of estimators: Simulation study}

In this section, a Monte Carlo simulation study is conducted for the purpose of evaluation and comparison of the finite-sample behavior of the maximum likelihood estimates, bias-corrected estimates obtained by using the Cox-Snell Methodology and the conditional mean estimates of the parameters of both the $ULZI$ and $ULZOI$ distributions.\\
Samples of size $n=25,30,50,100,500$ and 1000 are generated and 1000 Monte Carlo replications is considered. The five sets of parameters of $ULZI$ distribution used for this numerical exercise are $\alpha=0.2, E\left( Y|y\in \left( 0,1\right) \right) =0.1$; $\alpha=0.2, E\left( Y|y\in \left( 0,1\right) \right) =0.4$; $\alpha=0.2, E\left( Y|y\in \left( 0,1\right) \right) =0.7$; $\alpha=0.5,E\left( Y|y\right.\\
\left.\in\left(0,1\right)\right) =0.1$ and $\alpha=0.5,  E\left( Y|y\in \left( 0,1\right)\right) =0.4$, which corresponds to $\theta$ values $7.0, 1.0, 0.14, 4.0$ and $0.25$ respectively. The four sets of parameters of ULZOI distribution used are $\alpha=0.3, E\left( Y|y\in \left( 0,1\right) \right) =0.4, p=0.3; \alpha=0.3, E\left( Y|y\in \left( 0,1\right) \right) =0.6, p=0.5; \alpha=0.5, E\left( Y|y\in \left( 0,1\right) \right)=0.4, p=0.3 $ and $\alpha=0.5, E\left( Y|y\in \left( 0,1\right) \right) =0.6, p=0.5$, which corresponds $\theta$ values 1.26, 0.56, 1.0 and 0.43 respectively. To simulate $n$ observations from $ULZI\left( \alpha,\theta\right)$  distribution, the following algorithm was implemented:\\

\noindent{\bf Algorithm to generate from $ULZI\left( \alpha,\theta\right)$.}\\

\noindent{\it Step 1. $n$ random numbers from $U\left( 0,1\right) $ were generated, say $U_i; i=1,2,...,\ldots n $\\
Step 2. If $U_i < \alpha$, then we assign $y_i=0$.\\
Step 3. If $U_i \geq \alpha$, then we draw a random number from the Lindley($\theta$) distribution, say $x_i$ and assign $y_i=\dfrac{x_i}{1+x_i}$.}\\

Observations are simulated from the $ULZOI\left( \alpha,\theta, p\right)$ distribution using the following algorithm:\\

\noindent{\bf Algorithm to generate from $ULZOI\left( \alpha,\theta, p\right)$.}\\

\noindent{\it Step 1. $n$ random numbers from $U\left( 0,1\right) $ were generated, say $U_i; i=1,2,...,\ldots n $\\
Step 2. If $U_i \leq \alpha p$, then we assign $y_i=0$.\\
3. If $U_i \leq \alpha$, then we assign $y_i=1$.\\
Step 3. Otherwise, we draw a random number from the Lindley($\theta$) distribution, say $x_i$ and assign $y_i=\dfrac{x_i}{1+x_i}$.}\\

The performance evaluation of the estimates was done based on the estimated bias and Root Mean Square Error (RMSE).Table \ref{Tab1} and Table \ref{Tab2} present the simulation results for the ULZI distribution and ULZOI distribution respectively.

\begin{table}[h!]
	\scriptsize
	\centering
	\caption{Simulation results for ULZI distribution for different set of values of $\theta$ and $\alpha$}\label{Tab1}
	
	\tabcolsep=3pt
	\begin{tabular}{|c|c|c|c|c|c|c|c|c|c|c|}
		\hline
		\multicolumn{11}{|c|}{$\alpha=0.2,\theta=7$} \\
		\hline
		  & n & \multicolumn{3}{c}{Mean} \vrule & \multicolumn{3}{c}{Bias} \vrule& \multicolumn{3}{c}{RMSE}\vrule\\
		\cline{3-5} \cline{6-8} \cline{9-11}
		\cline{3-11}
		& &  MLE & BCMLE & CME & MLE & BCMLE & CME & MLE & BCMLE & CME\\
		\hline
		& 25 & 7.298 & 6.988 & 7.755 & 0.892 & -0.011 & 0.755 & 0.667 & 0.590 & 0.663\\
		& 50 & 7.165 & 7.007 & 7.130 & 0.165 & 0.007 & 0.130 & 0.110 & 0.108 & 0.112\\
		& 100 & 7.093 & 7.015 & 7.075 & 0.093 & 0.015 & 0.075 & 0.057 & 0.057 & 0.058\\
		$\theta$& 200 & 7.045 & 7.006 & 7.034 & 0.045 & 0.006 & 0.034 & 0.028 & 0.028 & 0.029\\
		& 500 & 7.029 & 7.013 & 7.024 & 0.028 & 0.013 & 0.024 & 0.011 & 0.011 & 0.011\\
		& 1000 & 7.023 & 7.016 & 7.018 & 0.024 & 0.016 & 0.018 & 0.006 & 0.006 & 0.006\\
		\hline
		  & n & \multicolumn{3}{c}{Mean} \vrule & \multicolumn{3}{c}{Bias} \vrule& \multicolumn{3}{c}{RMSE}\vrule\\
		\cline{3-5} \cline{6-8} \cline{9-11}
		\cline{3-11}
		& &  \multicolumn{3}{|c|}{MLE} & \multicolumn{3}{|c|}{MLE} & \multicolumn{3}{|c|}{MLE}\\
		\hline
		& 25 & \multicolumn{3}{|c|}{0.198} & \multicolumn{3}{|c|}{-0.001} & \multicolumn{3}{|c|}{0.012}\\
		& 50 & \multicolumn{3}{|c|}{0.202} & \multicolumn{3}{|c|}{0.003} & \multicolumn{3}{|c|}{0.006}\\
		& 100 & \multicolumn{3}{|c|}{0.201} & \multicolumn{3}{|c|}{0.001} & \multicolumn{3}{|c|}{0.003}\\
		$\alpha$ & 200 & \multicolumn{3}{|c|}{0.201} & \multicolumn{3}{|c|}{0.001} & \multicolumn{3}{|c|}{0.001}\\
		& 500 & \multicolumn{3}{|c|}{0.200} & \multicolumn{3}{|c|}{0.000} & \multicolumn{3}{|c|}{0.000}\\
		& 1000 & \multicolumn{3}{|c|}{0.199} & \multicolumn{3}{|c|}{0.000} & \multicolumn{3}{|c|}{0.000}\\
		\hline
		\hline
		\multicolumn{11}{|c|}{$\alpha=0.5,\theta=0.25$} \\
		\hline
		  & n & \multicolumn{3}{c}{Mean} \vrule & \multicolumn{3}{c}{Bias} \vrule& \multicolumn{3}{c}{RMSE}\vrule\\
		\cline{3-5} \cline{6-8} \cline{9-11}
		\cline{3-11}
		& &  MLE & BCMLE & CME & MLE & BCMLE & CME & MLE & BCMLE & CME\\
		\hline
		& 25 & 0.262 & 0.250 & 0.254 & 0.012 & 0.000 & 0.004 & 0.009 & 0.009 & 0.012\\
		& 50 & 0.256 & 0.251 & 0.253 & 0.006 & 0.001 & 0.003 & 0.004 & 0.004 & 0.006\\
		& 100 & 0.252 & 0.250 & 0.250 & 0.002 & -0.000 & 0.000 & 0.002 & 0.003 & 0.003\\
		$\theta$& 200 & 0.252 & 0.251 & 0.251 & 0.002 & 0.001 & 0.001 & 0.001 & 0.001 & 0.001\\
		& 500 & 0.250 & 0.250 & 0.250 & 0.000 & 0.000 & 0.000 & 0.000 & 0.000 & 0.000\\
		& 1000 & 0.250 & 0.250 & 0.250 & 0.000 & 0.000 & 0.000 & 0.000 & 0.000 & 0.000\\
		\hline
		  & n & \multicolumn{3}{c}{Mean} \vrule & \multicolumn{3}{c}{Bias} \vrule& \multicolumn{3}{c}{RMSE}\vrule\\
		\cline{3-5} \cline{6-8} \cline{9-11}
		\cline{3-11}
		& &  \multicolumn{3}{|c|}{MLE} & \multicolumn{3}{|c|}{MLE} & \multicolumn{3}{|c|}{MLE}\\
		\hline
		& 25 & \multicolumn{3}{|c|}{0.501} & \multicolumn{3}{|c|}{0.003} & \multicolumn{3}{|c|}{0.039}\\
		& 50 & \multicolumn{3}{|c|}{0.498} & \multicolumn{3}{|c|}{-0.002} & \multicolumn{3}{|c|}{0.008}\\
		& 100 & \multicolumn{3}{|c|}{0.499} & \multicolumn{3}{|c|}{-0.001} & \multicolumn{3}{|c|}{0.004}\\
		$\alpha$ & 200 & \multicolumn{3}{|c|}{0.499} & \multicolumn{3}{|c|}{-0.001} & \multicolumn{3}{|c|}{0.002}\\
		& 500 & \multicolumn{3}{|c|}{0.500} & \multicolumn{3}{|c|}{0.000} & \multicolumn{3}{|c|}{0.001}\\
		& 1000 & \multicolumn{3}{|c|}{0.499} & \multicolumn{3}{|c|}{-0.000} & \multicolumn{3}{|c|}{0.000}\\
		\hline 
	\end{tabular}
\end{table}

\begin{table}[h!]
	\scriptsize
	\centering
	\caption{Simulation results for ULZOI distribution for different set of values of $\theta$, $\alpha$ and $p$}\label{Tab2}
	\tabcolsep=3pt
	\begin{tabular}{|c|c|c|c|c|c|c|c|c|c|c|}
		\hline
		\multicolumn{11}{|c|}{$\alpha=0.3, p=0.5, \theta=0.56$} \\
		\hline
		  & n & \multicolumn{3}{c}{Mean} \vrule & \multicolumn{3}{c}{Bias} \vrule& \multicolumn{3}{c}{RMSE}\vrule\\
		\cline{3-5} \cline{6-8} \cline{9-11}
		\cline{3-11}
		& &  MLE & BCMLE & CME & MLE & BCMLE & CME & MLE & BCMLE & CME\\
		\hline
		& 30 & 0.571 & 0.556 & 0.567 & 0.011 & -0.004 & 0.007 & 0.013 & 0.013 & 0.016\\
		& 50 & 0.565 & 0.556 & 0.563 & 0.005 & -0.004 & 0.003 & 0.008 & 0.008 & 0.010\\
		& 100 & 0.563 & 0.559 & 0.562 & 0.003 & -0.001 & 0.002 & 0.004 & 0.004 & 0.005\\
		$\theta$& 200 & 0.562 & 0.559 & 0.561 & 0.002 & -0.001 & 0.001 & 0.002 & 0.002 & 0.002\\
		& 500 & 0.562 & 0.561 & 0.561 & 0.002 & 0.001 & 0.001 & 0.001 & 0.001 & 0.001\\
		& 1000 & 0.561 & 0.561 & 0.561 & 0.001 & 0.001 & 0.001 & 0.000 & 0.000 & 0.000\\
		\hline
		  & n & \multicolumn{3}{c}{Mean} \vrule & \multicolumn{3}{c}{Bias} \vrule& \multicolumn{3}{c}{RMSE}\vrule\\
		\cline{3-5} \cline{6-8} \cline{9-11}
		\cline{3-11}
		& &  \multicolumn{3}{|c|}{MLE} & \multicolumn{3}{|c|}{MLE} & \multicolumn{3}{|c|}{MLE}\\
		\hline
		& 30 & \multicolumn{3}{|c|}{0.301} & \multicolumn{3}{|c|}{0.001} & \multicolumn{3}{|c|}{0.012}\\
		& 50 & \multicolumn{3}{|c|}{0.296} & \multicolumn{3}{|c|}{-0.003} & \multicolumn{3}{|c|}{0.007}\\
		& 100 & \multicolumn{3}{|c|}{0.301} & \multicolumn{3}{|c|}{0.001} & \multicolumn{3}{|c|}{0.004}\\
		$\alpha$ & 200 & \multicolumn{3}{|c|}{0.300} & \multicolumn{3}{|c|}{-0.000} & \multicolumn{3}{|c|}{0.002}\\
		& 500 & \multicolumn{3}{|c|}{0.300} & \multicolumn{3}{|c|}{0.000} & \multicolumn{3}{|c|}{0.001}\\
		& 1000 & \multicolumn{3}{|c|}{0.300} & \multicolumn{3}{|c|}{0.000} & \multicolumn{3}{|c|}{0.000}\\
		\hline
		  & n & \multicolumn{3}{c}{Mean} \vrule & \multicolumn{3}{c}{Bias} \vrule& \multicolumn{3}{c}{RMSE}\vrule\\
		\cline{3-5} \cline{6-8} \cline{9-11}
		\cline{3-11}
		& &  \multicolumn{3}{|c|}{MLE} & \multicolumn{3}{|c|}{MLE} & \multicolumn{3}{|c|}{MLE}\\
		\hline
		& 30 & \multicolumn{3}{|c|}{0.504} & \multicolumn{3}{|c|}{0.004} & \multicolumn{3}{|c|}{0.025}\\
		& 50 & \multicolumn{3}{|c|}{0.506} & \multicolumn{3}{|c|}{0.006} & \multicolumn{3}{|c|}{0.014}\\
		& 100 & \multicolumn{3}{|c|}{0.502} & \multicolumn{3}{|c|}{0.002} & \multicolumn{3}{|c|}{0.007}\\
		$p$ & 200 & \multicolumn{3}{|c|}{0.501} & \multicolumn{3}{|c|}{0.001} & \multicolumn{3}{|c|}{0.004}\\
		& 500 & \multicolumn{3}{|c|}{0.498} & \multicolumn{3}{|c|}{-0.002} & \multicolumn{3}{|c|}{0.002}\\
		& 1000 & \multicolumn{3}{|c|}{0.500} & \multicolumn{3}{|c|}{-0.000} & \multicolumn{3}{|c|}{0.001}\\
		\hline
		\multicolumn{11}{|c|}{$\alpha=0.5, p=0.3, \theta=1$} \\
		\hline
		  & n & \multicolumn{3}{c}{Mean} \vrule & \multicolumn{3}{c}{Bias} \vrule& \multicolumn{3}{c}{RMSE}\vrule\\
		\cline{3-5} \cline{6-8} \cline{9-11}
		\cline{3-11}
		& &  MLE & BCMLE & CME & MLE & BCMLE & CME & MLE & BCMLE & CME\\
		\hline
		& 30 & 1.039 & 0.994 & 1.023 & 0.039 & -0.006 & 0.023 & 0.029 & 0.028 & 0.034\\
		& 50 & 1.025 & 0.999 & 1.015 & 0.025 & -0.001 & 0.015 & 0.017 & 0.017 & 0.015\\
		& 100 & 1.009 & 0.996 & 1.003 & 0.009 & -0.004 & 0.003 & 0.008 & 0.008 & 0.010\\
		$\theta$& 200 & 1.005 & 0.999 & 1.003 & 0.005 & -0.001 & 0.003 & 0.004 & 0.004 & 0.005\\
		& 500 & 1.003 & 1.001 & 1.003 & 0.003 & 0.001 & 0.003 & 0.002 & 0.002 & 0.002\\
		& 1000 & 1.001 & 1.000 & 1.001 & 0.002 & 0.000 & 0.001 & 0.001 & 0.001 & 0.001\\
		\hline
		  & n & \multicolumn{3}{c}{Mean} \vrule & \multicolumn{3}{c}{Bias} \vrule& \multicolumn{3}{c}{RMSE}\vrule\\
		\cline{3-5} \cline{6-8} \cline{9-11}
		\cline{3-11}
		& &  \multicolumn{3}{|c|}{MLE} & \multicolumn{3}{|c|}{MLE} & \multicolumn{3}{|c|}{MLE}\\
		\hline
		& 30 & \multicolumn{3}{|c|}{0.501} & \multicolumn{3}{|c|}{0.001} & \multicolumn{3}{|c|}{0.013}\\
		& 50 & \multicolumn{3}{|c|}{0.503} & \multicolumn{3}{|c|}{0.003} & \multicolumn{3}{|c|}{0.008}\\
		& 100 & \multicolumn{3}{|c|}{0.501} & \multicolumn{3}{|c|}{0.001} & \multicolumn{3}{|c|}{0.004}\\
		$\alpha$ & 200 & \multicolumn{3}{|c|}{0.500} & \multicolumn{3}{|c|}{0.000} & \multicolumn{3}{|c|}{0.002}\\
		& 500 & \multicolumn{3}{|c|}{0.501} & \multicolumn{3}{|c|}{0.001} & \multicolumn{3}{|c|}{0.001}\\
		& 1000 & \multicolumn{3}{|c|}{0.500} & \multicolumn{3}{|c|}{0.001} & \multicolumn{3}{|c|}{0.000}\\
		\hline
		  & n & \multicolumn{3}{c}{Mean} \vrule & \multicolumn{3}{c}{Bias} \vrule& \multicolumn{3}{c}{RMSE}\vrule\\
		\cline{3-5} \cline{6-8} \cline{9-11}
		\cline{3-11}
		& &  \multicolumn{3}{|c|}{MLE} & \multicolumn{3}{|c|}{MLE} & \multicolumn{3}{|c|}{MLE}\\
		\hline
		& 30 & \multicolumn{3}{|c|}{0.305} & \multicolumn{3}{|c|}{0.005} & \multicolumn{3}{|c|}{0.017}\\
		& 50 & \multicolumn{3}{|c|}{0.298} & \multicolumn{3}{|c|}{-0.002} & \multicolumn{3}{|c|}{0.011}\\
		& 100 & \multicolumn{3}{|c|}{0.299} & \multicolumn{3}{|c|}{-0.001} & \multicolumn{3}{|c|}{0.005}\\
		$p$ & 200 & \multicolumn{3}{|c|}{0.300} & \multicolumn{3}{|c|}{-0.000} & \multicolumn{3}{|c|}{0.003}\\
		& 500 & \multicolumn{3}{|c|}{0.301} & \multicolumn{3}{|c|}{0.002} & \multicolumn{3}{|c|}{0.001}\\
		& 1000 & \multicolumn{3}{|c|}{0.301} & \multicolumn{3}{|c|}{0.001} & \multicolumn{3}{|c|}{0.001}\\
		\hline
		\end{tabular}
\end{table}
\newpage
\noindent Table 1 shows that for the ULZI distribution, the bias corrected estimate of both $\theta$ achieves substantial bias reduction over the conditional mean estimate and maximum likelihood estimate whereas the RMSE of MLE and BCMLE are smaller than those of CME. Both the bias and RMSE decreases with an increase in $n$. For moderately large and large sample sizes, the bias of $\alpha$ is seen to be negative and the RMSE of $\alpha$ also decreases with an increase in $n$.\\
From Table 2, it is evident that for the ULZOI distribution, for small and moderately large sample sizes, the bias of BCMLE of $\theta$ is negative and the RMSE of both MLE and BCMLE coincide and are less than that of the CME. The RMSE of the estimates of $\theta$, $\alpha$ and $p$ decrease with an increase in $n$. Further, the bias of $p$ is seen to be negative for moderately large and large sample sizes.   
Table 3 and 4 display the confidence intervals and empirical coverage probabilities at 90\% and 95\% for the sample sizes $n=100, 200, 500$ and 1000 for the ULZI and ULZOI distribution respectively.
\begin{table}[h!]
	\scriptsize
	\centering
	\caption{Confidence intervals and coverage probabilities for ULZI distribution for different set of values of $\theta$ and $\alpha$}\label{Tab3}
	\tabcolsep=2pt
	\begin{tabular}{|c|c|c|c|c|c|c|c|c|c|}
		\hline
		\multicolumn{10}{|c|}{$\alpha=0.2,\theta=7$} \\
		\hline
		  & $\% $ & \multicolumn{4}{c}{Confidence interval} \vrule & \multicolumn{4}{c}{Coverage probability}\vrule\\
		\cline{3-6} \cline{7-10}
		& &  n=50 & n=100 & n=500 & n=1000 & n=50 & n=100 & n=500 & n=1000\\
		\hline
		$\theta$ & 95\% & (5.047,8.952)	& (5.619,8.381) & (6.383,7.617) & (6.563,7.436) & 94.4 & 94 & 95.2 & 94.5\\
		\cline{2-10}
		& 90\% & (5.361,8.638) & (5.841,8.159) & (6.481,7.518) & (6.634,7.366) & 90.8 & 89.6 & 91.6 & 88.7\\
		\hline
		$\alpha$ & 95\% & (0.089,0.311)	& (0.122,0.278) & (0.165,0.235) & (0.175,0.225) & 95.1 & 93.7 & 95.2 & 94.4\\
		\cline{2-10}
		& 90\% & (0.107,0.293) & (0.134,0.266) & (0.171,0.229) & (0.179,0.221) & 88.5 & 88.9 & 91.6 & 88.9\\
		\hline
		\multicolumn{10}{|c|}{$\alpha=0.5,\theta=0.25$} \\
		\hline
	  & $ \% $ & \multicolumn{4}{c}{Confidence interval} \vrule & \multicolumn{4}{c}{Coverage probability}\vrule\\
		\cline{3-6} \cline{7-10}
		& &  n=50 & n=100 & n=500 & n=1000 & n=50 & n=100 & n=500 & n=1000\\
		\hline
		$\theta$ & 95\% & (0.207,0.293)	& (0.219,0.281) & (0.236,0.264) & (0.240,0.260) & 76.9 & 77.8 & 76.9 & 79.6\\
		\cline{2-10}
		& 90\% & (0.214,0.286) & (0.224,0.276) & (0.238,0.262) & (0.242,0.258) & 69.7 & 67.9 & 69.6 & 71.4\\
		\hline
		$\alpha$ & 95\% & (0.361,0.639)	& (0.402,0.598) & (0.456,0.543) & (0.469,0.531) & 92.8 & 95 & 92.9 & 94.6\\
		\cline{2-10}
		& 90\% & (0.384,0.616) & (0.417,0.582) & (0.463,0.538) & (0.474,0.526) & 86.5 & 92 & 88.7 & 89.8\\
		\hline			
	\end{tabular}
\end{table}

\begin{table}[h!]
	\scriptsize
	\centering
	\caption{Confidence intervals and coverage probabilities for ULZOI distribution for different set of values of $\theta$, $\alpha$ and $p$}\label{Tab4}
	\tabcolsep=2pt
	\begin{tabular}{|c|c|c|c|c|c|c|c|c|c|}
		\hline
		\multicolumn{10}{|c|}{$\alpha=0.3,p=0.5,\theta=0.56$} \\
		\hline
		  & $\% $ & \multicolumn{4}{c}{Confidence interval} \vrule & \multicolumn{4}{c}{Coverage probability}\vrule\\
		\cline{3-6} \cline{7-10}
		& &  n=50 & n=100 & n=500 & n=1000 & n=50 & n=100 & n=500 & n=1000\\
		\hline
		$\theta$ & 95\% & (0.485,0.635)	& (0.507,0.613) & (0.562,0.594) & (0.536,0.584) & 86.9 & 88.8 & 87.9 & 86.2\\
		\cline{2-10}
		& 90\% & (0.497,0.623) & (0.515,0.605) & (0.532,0.588) & (0.540,0.580) & 79.8 & 80.8 & 81.7 & 79.3\\
		\hline
		$\alpha$ & 95\% & (0.210,0.390)	& (0.236,0.364) & (0.260,0.340) & (0.272,0.328) & 93.2 & 95.1 & 95.6 & 94.2\\
		\cline{2-10}
		& 90\% & (0.224,0.375) & (0.247,0.353) & (0.266,0.334) & (0.276,0.324) & 89 & 90 & 89.5 & 88\\
		\hline
		$p$ & 95\% & (0.321,0.679)	& (0.373,0.627) & (0.420,0.580) & (0.443,0.557) & 95.1 & 94.4 & 94.4 & 95.1\\
		\cline{2-10}
		& 90\% & (0.344,0.650) & (0.394,0.606) & (0.432,0.567) & (0.453,0.547) & 89.9 & 88.4 & 88.7 & 89.7\\
		\hline
		\multicolumn{10}{|c|}{$\alpha=0.5,p=0.3,\theta=1$} \\
		\hline
		  & $\% $ & \multicolumn{4}{c}{Confidence interval} \vrule & \multicolumn{4}{c}{Coverage probability}\vrule\\
		\cline{3-6} \cline{7-10}
		& &  n=50 & n=100 & n=500 & n=1000 & n=50 & n=100 & n=500 & n=1000\\
		\hline
		$\theta$ & 95\% & (0.846,1.154)	& (0.891,1.109) & (0.931,1.069) & (0.951,1.049) & 85.4 & 85.4 & 85.4 & 83.7\\
		\cline{2-10}
		& 90\% & (0.871,1.129) & (0.909,1.091) & (0.942,1.058) & (0.959,1.041) & 79.7 & 78.9 & 77.1 & 75.6\\
		\hline
		$\alpha$ & 95\% & (0.402,0.598)	& (0.431,0.569) & (0.456,0.544) & (0.469,0.531) & 95 & 93.2 & 95.1 & 93.8\\
		\cline{2-10}
		& 90\% & (0.418,0.582) & (0.442,0.558) & (0.463,0.538) & (0.474,0.526) & 91.5 & 88.7 & 90 & 90.9\\
		\hline	
		$p$ & 95\% & (0.173,0.427)	& (0.210,0.390) & (0.243,0.357) & (0.260,0.340) & 95.1 & 95.5 & 93.9 & 94.7\\
		\cline{2-10}
		& 90\% & (0.193,0.407) & (0.225,0.375) & (0.252,0.348) & (0.266,0.334) & 90.6 & 90.1 & 89 & 90.1\\
		\hline		
	\end{tabular}
\end{table}
\noindent It can be seen from Table 3 that the accuracy of the empirical confidence intervals increases with an increase in the sample size for the single inflation case for both the parameters. The coverage probabilities of both $\alpha$ and $\theta$ are also seen to be above 85\% for all sample sizes when the value of $\theta$ is high, i.e. when $\theta=7$, whereas the coverage probability of $\theta$ is seen to be low (less then 80\%) when true value of $\theta$ is low, i.e. $\theta=0.25$.

\section{Applications} 
\begin{figure}[h!]
	\includegraphics[keepaspectratio,height=13cm,width=13cm]{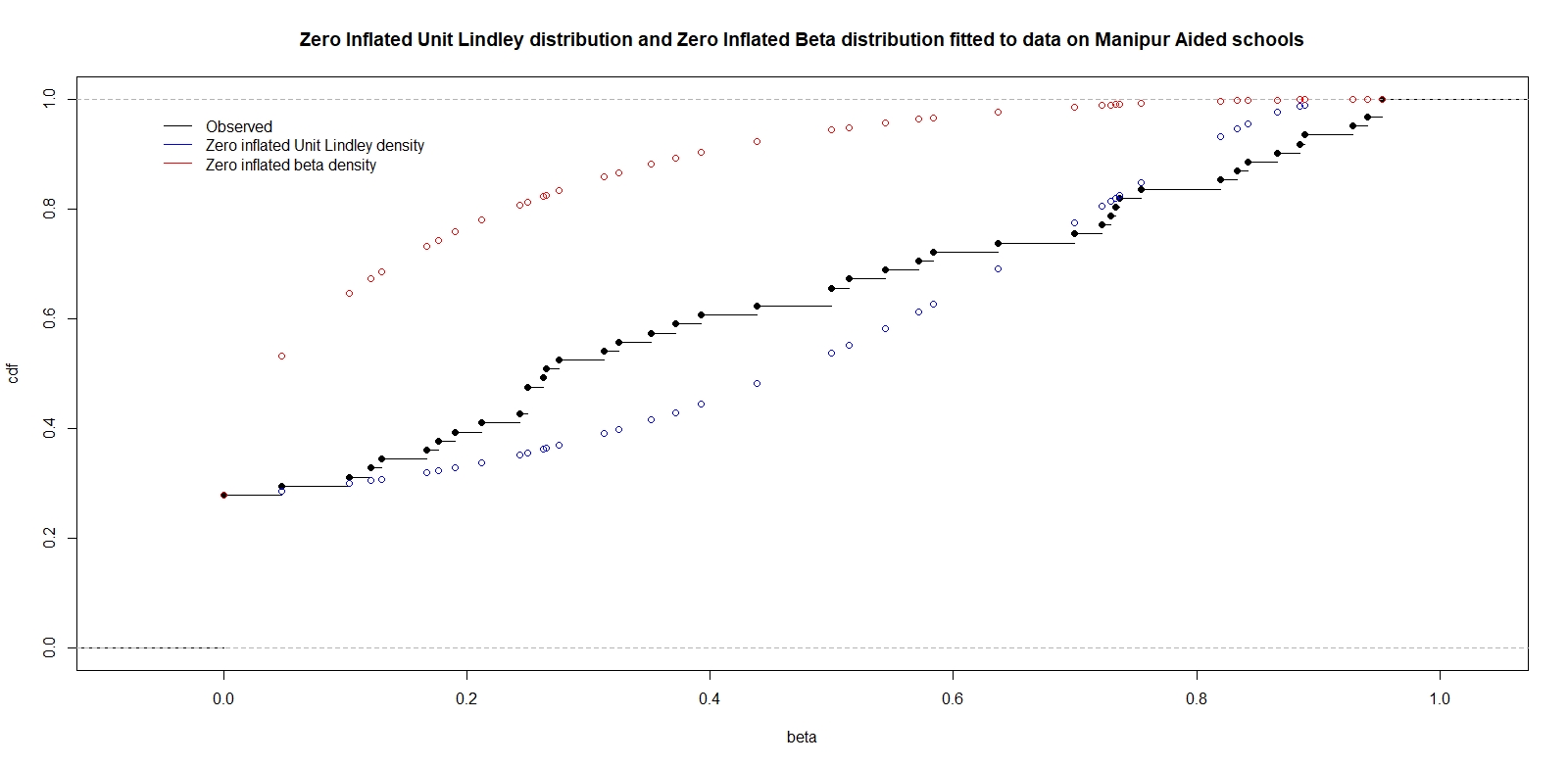}
	\caption{Observed d.f and c.d.f. plot of the fitted ULZI and ZIB distributions for proportion of students passing the H.S.L.C. exam 2020, Manipur from aided schools}\label{aided}
\end{figure}

\begin{figure}[h!]
	\includegraphics[keepaspectratio,height=13cm,width=13cm]{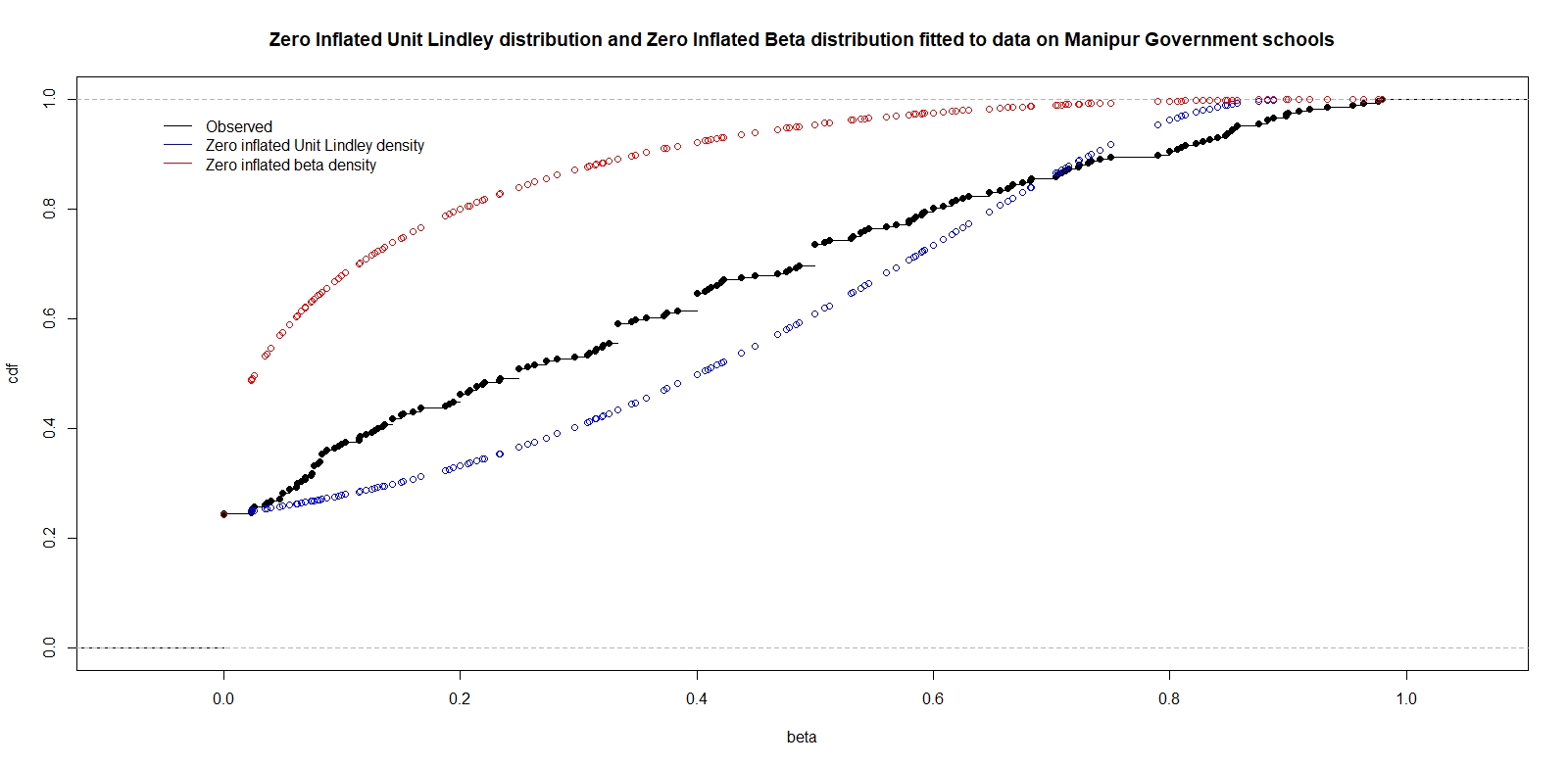}
	\caption{Observed d.f and c.d.f. plot of the fitted ULZI and ZIB distributions for proportion of students passing the H.S.L.C. exam 2020, Manipur from government schools}\label{govt}
\end{figure}

\begin{figure}[h!]
	\includegraphics[keepaspectratio,height=13cm,width=13cm]{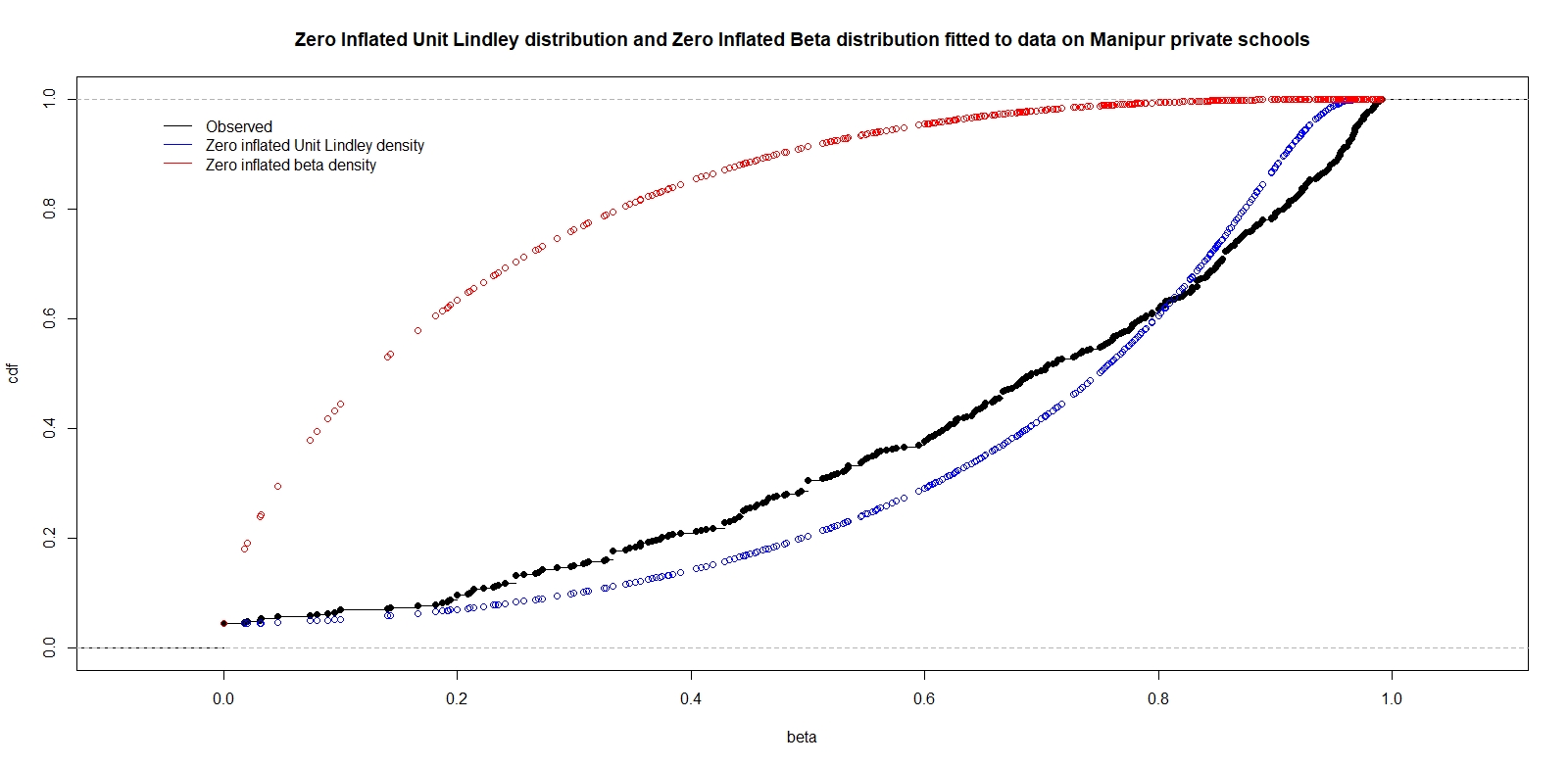}
	\caption{Observed d.f and c.d.f. plot of the fitted ULZI and ZIB distributions for proportion of students passing the H.S.L.C. exam 2020, Manipur from private schools}\label{pvt}
\end{figure}

\begin{figure}[h!]
	\includegraphics[keepaspectratio,height=13cm,width=13cm]{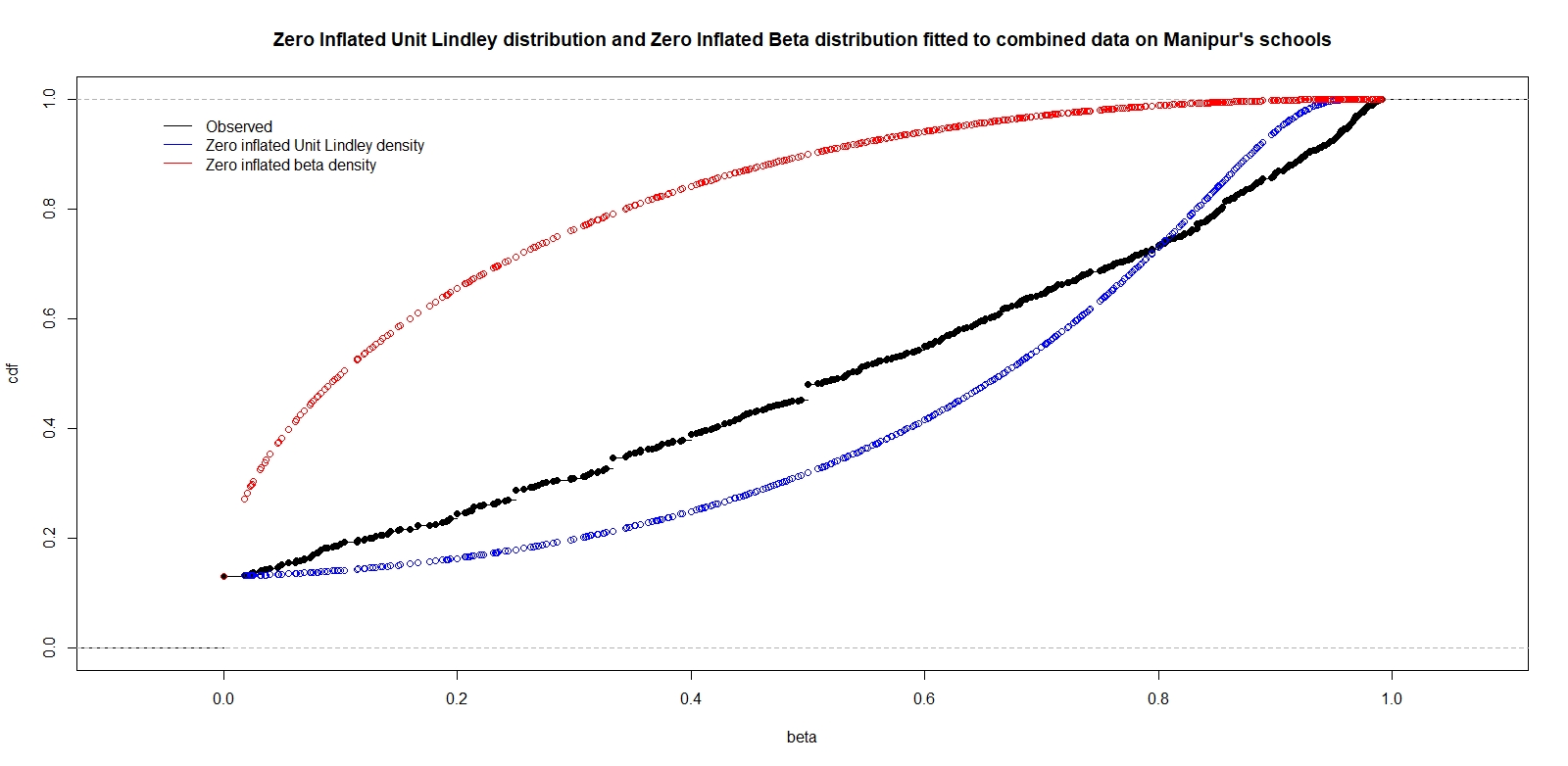}
	\caption{Observed d.f and c.d.f. plot of the fitted ULZI and ZIB distributions for proportion of students passing the H.S.L.C. exam 2020, Manipur from all schools combined}\label{comb}
\end{figure}

\begin{figure}[h!]
	\includegraphics[keepaspectratio,height=13cm,width=13cm]{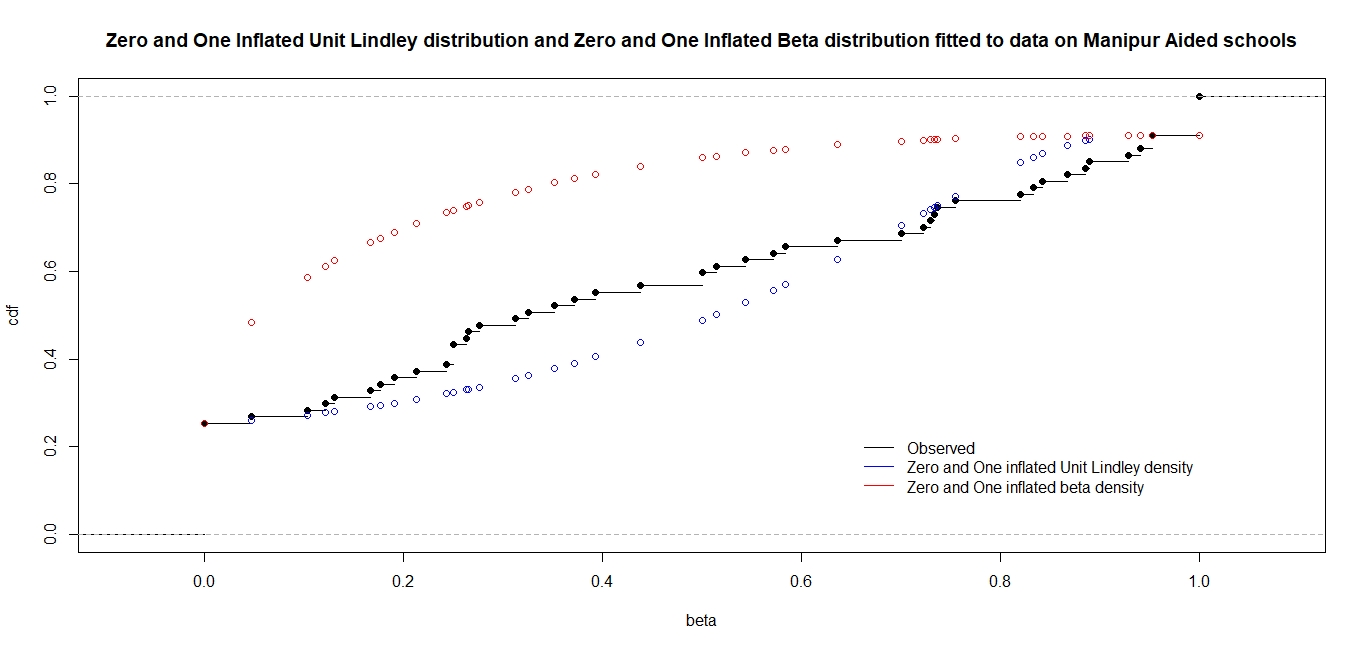}
	\caption{Observed d.f and c.d.f. plot of the fitted ULZOI and ZOIB distributions for proportion of students passing the H.S.L.C. exam 2020, Manipur from aided schools}\label{aided2}
\end{figure}

\begin{figure}[h!]
	\includegraphics[keepaspectratio,height=13cm,width=13cm]{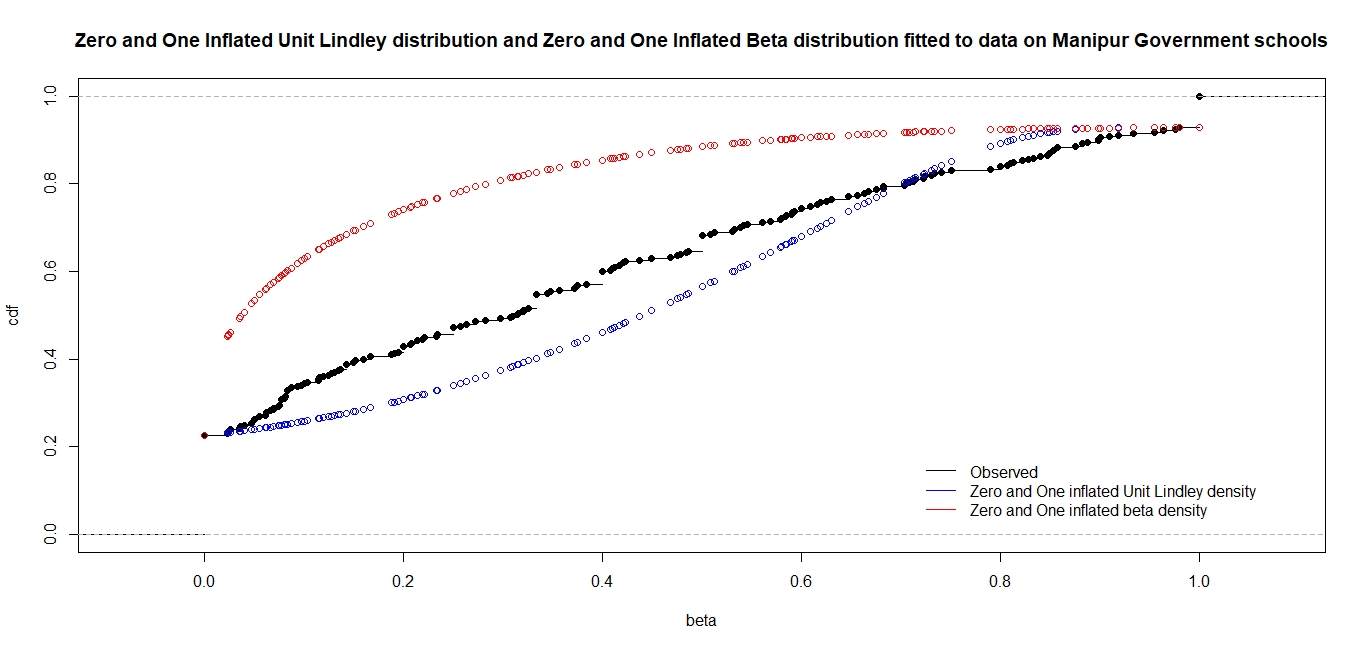}
	\caption{Observed d.f and c.d.f. plot of the fitted ULZOI and ZOIB distributions for proportion of students passing the H.S.L.C. exam 2020, Manipur from government schools}\label{govt2}
\end{figure}

\begin{figure}[h!]
	\includegraphics[keepaspectratio,height=13cm,width=13cm]{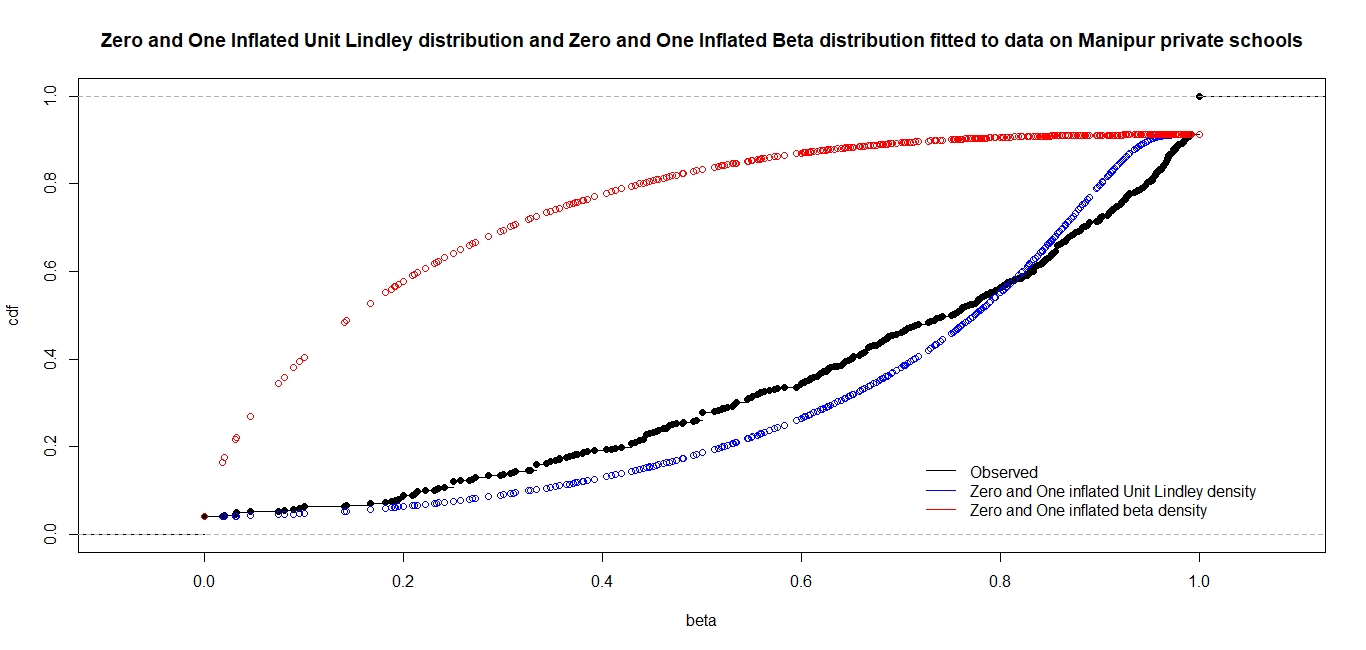}
	\caption{Observed d.f and c.d.f. plot of the fitted ULZOI and ZOIB distributions for proportion of students passing the H.S.L.C. exam 2020, Manipur from private schools}\label{pvt2}
\end{figure}

\begin{figure}[h!]
	\includegraphics[keepaspectratio,height=13cm,width=13cm]{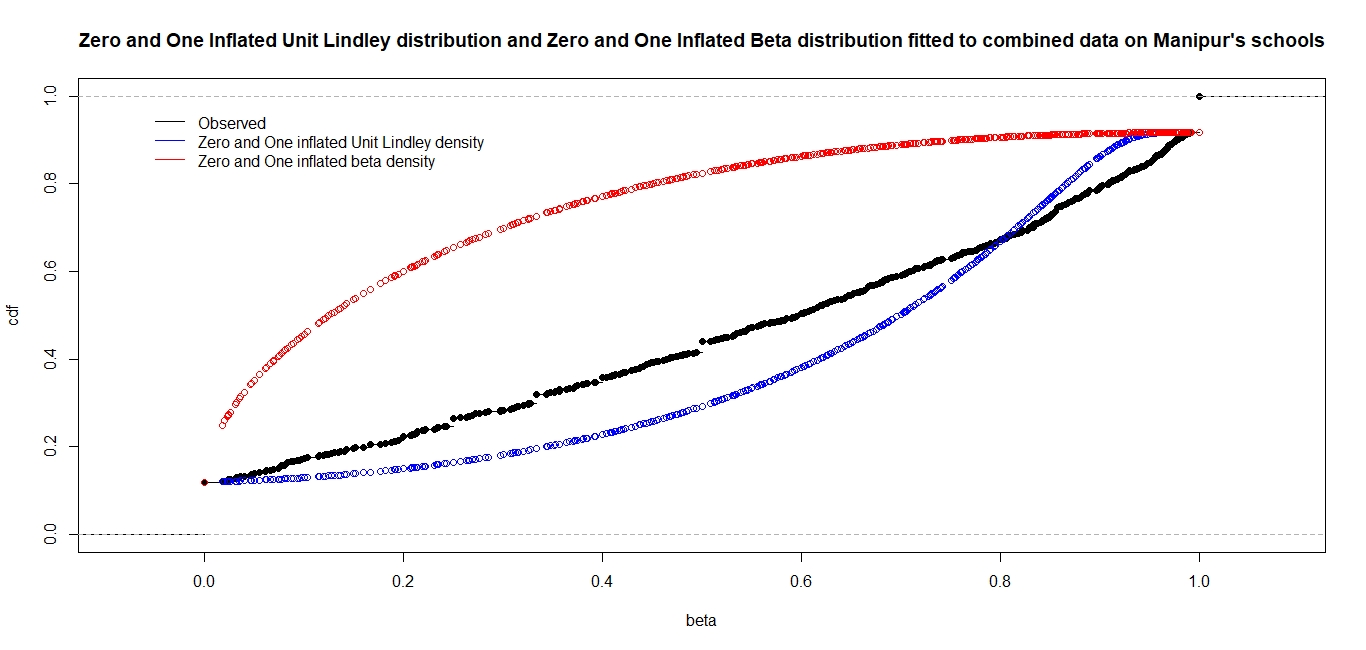}
	\caption{Observed d.f and c.d.f. plot of the fitted ULZOI and ZOIB distributions for proportion of students passing the H.S.L.C. exam 2020, Manipur from all schools combined}\label{comb2}
\end{figure}

In this section we consider real life data arising from High School Leaving Examination results of the State of Manipur, in India for the year 2020 (\cite{Boa202}, \cite{Boa201},\cite{Boa203}).  In both the cases, the variable of interest is the proportion of students who have passed the exam and it takes values in the interval $\left[ 0,1\right] $. The data have been collected from the website E-pao, an e-platform for Manipuris and it consists of 523 observations pertaining to private schools, 305 observations pertaining to government schools and 67 observations pertaining to aided schools. We compare our model with the famed inflated beta distribution \cite{Osp10}. \\
Table \ref{Tab5} displays the descriptive statistics of the data sets on proportion of students studying in government schools, private schools and aided schools of Manipur, India who have passed the High School Leaving Examination, 2020:
\begin{table}[h!]
\scriptsize
	\centering
	\caption{Descriptive statistics}\label{Tab5}
	\tabcolsep=2pt
	\begin{tabular}{|c|c|c|c|c|c|c|}
		\hline
		Data set & Minimum & Maximum & Mean & 1st quartile & Median & 3rd quartile\\
		\hline
		Aided schools & 0 & 1 & 0.4219 & 0.0238 & 0.325 & 0.7456\\
		Government schools & 0 & 1 & 0.3632 & 0.0476 & 0.3143 & 0.6154\\
		Private schools & 0 & 1 & 0.6705 & 0.4706 & 0.752 & 0.9189\\
		All schools combined & 0 & 1 & 0.5471 & 0.25 & 0.6 & 0.8621\\
		\hline
	\end{tabular}
\end{table}
\newline
We see from Table \ref{Tab5} that 75\% of the data points in the data set pertaining to aided schools, government schools, private schools and all schools combined exceed 0.0238, 0.0476, 0.4706 and 0.5471 respectively. Further, the mean exceeds the median for aided and government schools. Thus, it is clear that the data sets on proportion of students studying in aided and government schools are right skewed. Similarly, the data sets on proportion of students studying in private schools and all schools combined are left skewed.\\
The Zero and Zero-and-One Inflated Unit Lindley distributions and Zero and Zero-and-One Inflated Beta distributions are fitted to each of the four data sets, where the estimation of parameters have been done using the maximum likelihood method. The Kolmogorov-Smirnov (KS) test statistic is used to compare the fit of each of the distributions to the data sets. Table \ref{Tab6} displays the maximum likelihood estimates and standard errors of the parameters and Kolmogorov-Smirnov test statistic values for the Zero Inflated Unit Lindley distribution and Zero Inflated Beta distribution fitted to the four data sets:

\begin{table}[h!]
	\scriptsize
	\centering
	\caption{ML estimates and standard errors of the parameters of and Kolmogorov-Smirnov test statistic values for ULZI and ZIB distributions fitted to the four data sets}\label{Tab6}
	\tabcolsep=1pt
	\begin{tabular}{|c|c|c|c|}
		\hline
		Data set & Distribution & MLE \& SE of the parameters & K-S test statistic value\\
		\hline
		\multirow{2}{*}{Aided schools} & ULZI & MLE: $\alpha=0.2787,\theta=0.5273$ & \\
		\cline{3-3}
		& & SE: $\alpha=0.0547,\theta=0.0459$ & 0.2357\\
		\cline{2-4}
		& ZIB & MLE: $\alpha=0.2787,\mu=0.5143,\phi=2.5075$ & \\
		\cline{3-3}
		& & SE: $\alpha=0.0547,\mu=0.0389,\phi=0.4595$ & 0.3809\\
		\hline
		\multirow{2}{*}{Government schools} & ULZI & MLE: $\alpha=0.2438,\theta=0.7617$ & \\
		\cline{3-3}
		& & SE: $\alpha=0.0255,\theta=0.0328$ & 0.2411\\
		\cline{2-4}
		& ZIB & MLE: $\alpha=0.2438,\mu=0.4207,\phi=2.5035$ & \\
		\cline{3-3}
		& & SE: $\alpha=0.0255,\mu=0.0174,\phi=0.2096$ & 0.3452\\
		\hline
		\multirow{2}{*}{Private schools} & ULZI & MLE: $\alpha=0.0440,\theta=0.2288$ & \\
		\cline{3-3}
		& & SE: $\alpha=0.0094,\theta=0.0065$ & 0.3148\\
		\cline{2-4}
		& ZIB & MLE: $\alpha=0.0440,\mu=0.6599,\phi=2.6921$ & \\
		\cline{3-3}
		& & SE: $\alpha=0.0094,\mu=0.0112,\phi=0.1591$ & 0.6466\\
		\hline
		\multirow{2}{*}{All schools combined} & ULZI & MLE: $\alpha=0.1303,\theta=0.3011$ & \\
		\cline{3-3}		
		& & SE: $\alpha=0.0117,\theta=0.0067$ & 0.3141\\
		\cline{2-4}
		& ZIB & MLE: $\alpha=0.1303,\mu=0.5772,\phi=2.1898$ & \\
		\cline{3-3}
		& & SE: $\alpha=0.0117,\mu=0.0099,\phi=0.0988$ & 0.4616\\
		\hline
	\end{tabular}
\end{table}
Figure \ref{aided}, \ref{govt}, \ref{pvt} and \ref{comb} show the plot of the observed distribution function and the distribution function of ULZI and ZIB distributions fitted to the four data sets. By visual inspection of the figures \ref{aided} to \ref{comb}, it is clear that the proposed Zero inflated Unit Lindley distribution is a better fit to all the four data sets in comparison to the popular Inflated Beta distribution.
From table \ref{Tab6}, we observe that for each of the four data sets, the value of the K-S test statistic for the ULZI distribution is less than that of the ZIB distribution. Thus, we conclude that the proposed Zero inflated Unit Lindley distribution is able to model each of the four dats sets better than the Zero inflated Beta distribution. 
\newline
Further, the Zero and One Inflated Unit Lindley distribution (ULZOI) and Zero and One Inflated Beta distribution (ZOIB) are fitted to each of the four data sets, with the estimation of parameters being carried out using the maximum likelihood method and the goodness of fit being tested using the Kolmogorov-Smirnov statistic. Table \ref{Tab7} exhibits the maximum likelihood estimates and standard errors of the parameters and Kolmogorov-Smirnov test statistic values for the ULZOI and ZOIB distributions fitted to the four data sets:

\begin{table}[h!]
	\scriptsize
	\centering
	\caption{ML estimates and standard errors of the parameters of and Kolmogorov-Smirnov test statistic values for ULZOI and ZOIB distributions fitted to the four data sets}\label{Tab7}
	\tabcolsep=1.5pt
	\begin{tabular}{|c|c|c|c|}
		\hline
		Data set & Distribution & MLE(SE) of the parameters & K-S test \\
		& & & statistic value\\
		\hline
		\multirow{2}{*}{Aided schools} & ULZOI & MLE: $\alpha=0.2022,\theta=0.3011,p=0.4088$ & \\
		\cline{3-3}
		& & SE: $\alpha=0.0132,\theta=0.0080,p=0.0361$ & 0.2881\\
		\cline{2-4}
		& ZOIB & $\alpha=0.2022,\mu=0.5772,\phi=2.1898,p=0.4088$ & \\
		\cline{3-3}
		& & SE: $\alpha=0.0132,\mu=0.0099,\phi=0.0988,p=0.0361$ & 0.3967\\
		\hline
		\multirow{2}{*}{Government schools} & ULZOI & MLE: $\alpha=0.2984,\theta=0.7617,p=0.2416$ & \\
		\cline{3-3}
		& & SE: $\alpha=0.0262,\theta=0.0387,p=0.0490$ & 0.2237\\
		\cline{2-4}
		& ZOIB & $\alpha=0.2984,\mu=0.4207,\phi=2.5035,p=0.2416$ & \\
		\cline{3-3}
		& & SE: $\alpha=0.0262,\mu=0.0174,\phi=0.2096,p=0.0490$ & 0.2885\\
		\hline
		\multirow{2}{*}{Private schools} & ULZOI & MLE: $\alpha=0.1281,\theta=0.2288,p=0.6866$ & \\
		\cline{3-3}
		& & SE: $\alpha=0.0143,\theta=0.0076,p=0.0581$ & 0.2870\\
		\cline{2-4}
		& ZOIB & MLE: $\alpha=0.1405,\mu=0.6599,\phi=2.6922,p=0.6866$ & \\
		\cline{3-3}
		& & SE: $\alpha=0.0143,\mu=0.0112,\phi=0.1593,p=0.0581$ & 0.5706\\
		\hline
		\multirow{2}{*}{All schools combined} & ULZOI & MLE: $\alpha=0.2022,\theta=0.3011,p=0.4088$ & \\
		\cline{3-3}
		& & SE: $\alpha=0.0132,\theta=0.0080,p=0.0361$ & 0.2881\\
		\cline{2-4}
		& ZOIB & MLE: $\alpha=0.2022,\mu=0.5772,\phi=2.1898,p=0.4088$ & \\
		\cline{3-3}
		& & SE: $\alpha0.0132,\mu=0.0099,\phi=0.0988,p=0.0361$ & 0.3967\\
		\hline
	\end{tabular}
\end{table}
Table \ref{Tab7} clearly shows that for each of the four data sets, the K-S test statistic value for the ULZOI distribution is smaller than that for the ZOIB distribution and so, the Zero and One Inlfated Unit Lindley distribution is able to model these data sets better than the Zero Inflated Beta distribution. Figure \ref{aided2}, \ref{govt2}, \ref{pvt2} and \ref{comb2} show the plot of the observed distribution function and the distribution function of ULZI and ZIB distributions fitted to the four data sets.
\newline
It is evident from figures \ref{aided2} to \ref{comb2} that the Zero and One Inflated Unit Lindley distribution is a better fit than the Zero and One Inflated Beta distribution to each of the four data sets. The same is confirmed by comparing the Kolmogorov-Smirnov test statistic values also. This is an obvious result as all the four data sets contain both zeroes and ones.
\newpage
\section{Concluding remarks}
Applied statisticians may often encounter data sets where the variable of interest assumes values in the interval $\left( 0,1\right) $, for instance, proportions, ratios, etc. However, the variable may also display the phenomenon of inflation, i.e., it may take the values zero and/or one with positive probabilities. In other words, the interest may lie in modeling variables assuming values in the intervals $\left[ 0,1\right) $, $\left( 0,1\right] $ or $\left[   0,1\right]   $. Inflated Beta distribution is the leading probability model existing in the current literature to model such variables. In the present work, the zero-or-one inflated version and zero-and-one inflated version of the unit Lindley distribution is discussed and its various important distributional aspects are studied. The finite sample behaviour of the point estimates and interval estimates are studied with the help of Monte Carlo simulation studies. Real life data modelling is performed on data sets of pass proportion of students to show the advantage of modelling them using the proposed models over the existing inflated versions of Beta distribution. Our findings confirmed the usefulness of the inflated version of unit Lindley over the considered data.

\end{document}